\pdfoutput=1

\documentclass[twocolumn]{aastex62}
\usepackage{amsmath,amstext}

\graphicspath{{./}{figures/}}

\received{January 1, 2018}
\revised{January 7, 2018}
\accepted{\today}
\submitjournal{ApJ}

\shorttitle{}
\shortauthors{Joseph et. al.}

\begin{document}

\title{The Bias and Uncertainty of Redundant and Sky-Based Calibration under Realistic Sky and Telescope Conditions}

\correspondingauthor{R.C. Joseph}
\email{ronniy.joseph@icrar.org}

\author[0000-0003-3457-4670]{Ronniy C. Joseph}
\affiliation{International Centre for Radio Astronomy Research - Curtin University\\
1 Turner Avenue, Bentley WA 6102, Australia}
\affiliation{ARC Centre of Excellence for All Sky Astrophysics in 3 Dimensions (ASTRO 3D), Perth, WA 6845}
\author{Cathryn M. Trott}
\affiliation{International Centre for Radio Astronomy Research - Curtin University\\
1 Turner Avenue, Bentley WA 6102, Australia}
\affiliation{ARC Centre of Excellence for All Sky Astrophysics in 3 Dimensions (ASTRO 3D), Perth, WA 6845}
\author{Randall B. Wayth}
\affiliation{International Centre for Radio Astronomy Research - Curtin University\\
1 Turner Avenue, Bentley WA 6102, Australia}
\affiliation{ARC Centre of Excellence for All Sky Astrophysics in 3 Dimensions (ASTRO 3D), Perth, WA 6845}



\begin{abstract}
The advent of a new generation of low frequency interferometers has opened a direct window into the Epoch of Reionisation (EoR). However, key to a detection of the faint 21-cm signal, and reaching the sensitivity limits of these arrays, is a detailed understanding of the instruments and their calibration. In this work we use simulations to investigate the bias and uncertainty of redundancy based calibration. Specifically, we study the influence of the flux distribution of the radio sky and the impact of antenna position offsets on the complex calibration solutions. We find that the position offsets introduce a bias into the phase component of the calibration solutions. This phase bias increases with the distance between bright radio sources and the pointing center, and with the flux density of these sources. This is potentially problematic for redundant calibration on MWA observations of EoR fields 1 and 2. EoR field 0, however, lacks such sources. We also compared the simulations with theoretical estimates for the bias and uncertainty in sky model based calibration on incomplete sky models for the redundant antenna tiles in the MWA. Our results indicate that redundant calibration outperforms sky based calibration due to the high positional precision of the MWA antenna tiles.       
\end{abstract}

\keywords{Astronomical Instrumentation, Methods and Techniques - early universe - instrumentation: interferometers - methods: numerical - techniques: interferometric }


\section{Introduction}
Over the past few years the latest generation of low-frequency interferometers has pushed down the upper limits of the 21-cm power spectrum of the Epoch of Reionisation (EoR) \citep{Beardsley2016, Patil2017}. However, none of the current instruments, e.g. the Murchison Widefield Array (MWA) \citep{Tingay2013}, the LOw Frequency ARray (LOFAR) \citep{vanHaarlem2013}, and the Precision Array for Probing the Epoch of Reionization (PAPER) \citep{Parsons2010}, have detected a signal thus far. The signal, emitted by neutral hydrogen during the EoR, is a direct probe into the state of the Intergalactic Medium (IGM) \citep{Furlanetto2006} and allows us to directly study the conditions under which the first luminous objects were formed. For more in depth reviews see \citet{Morales2010,Pritchard2012,McQuinn2015,Furlanetto2016}.

Foreground sources and instrumental effects pose large challenges to the detection of this faint signal. The low frequency foregrounds, e.g. the Milky Way and extragalactic sources, are expected to be 4-5 orders of magnitude stronger than the neutral hydrogen signal \citep{Furlanetto2006,Bowman2006,Morales2006,Pritchard2008,Jelic2008}. The removal of these foregrounds requires a detailed understanding of the instrument and its calibration, because the subtraction of these foregrounds, in particular bright compact sources, is sensitive to calibration errors \citep{Datta2009}. Failing to remove bright sources accurately leads to contamination of EoR data, causing the so-called ``wedge'' feature in the 2D-Power Spectrum. \citep{Datta2010,Morales2012,Trott2012,Vedantham2012} This leakage of bright source residuals into the power spectrum, makes certain scales of the EoR signal inaccessible if not dealt with correctly.

Adequate removal of these foregrounds and extraction of the faint signal from the data puts stringent requirements on our calibration accuracy and precision. Standard calibration schemes correct the sky signal cross-correlations (or `visibilities') measured by radio interferometers using sky models, hereby solving for the gain factors that cause the discrepancy between the modelled visibilities and the measured visibilities \citep[see ][for a review]{Rau2009}. Sky model based calibration has undergone tremendous progress in the past years in order to overcome direction dependent calibration effects, e.g. varying antenna primary beam shapes, and ionospheric distortions, that limit this new generation of instruments. This progress resulted in a large variety of improved sky based calibration implementations, e.g. \texttt{RTS} \citep{Mitchell2008}, \texttt{SAGEcal} \citep{Yatawatta2009, Kazemi2011}, \texttt{SPAM} \citep{Intema2009}, \texttt{FHD} \citep{Sullivan2012},  and facet calibration \citep{Weeren2016} to name a few. Nevertheless, at the operating frequency of these new low-frequency interferometers (80--200 MHz), our limited understanding of the sky leads to incomplete models. Model-based calibration with incomplete sky models causes calibration errors that lead to image artifacts, which in turn limit the dynamic range of observations \citep{Grobler2014,Wijnholds2016} and, more relevant to EoR science, contaminate the power spectrum \citep{Barry2016, Ewall-Wice2016, Trott2017}.
Redundant calibration, however, allows us to escape our ignorance of the low frequency sky because it does not require modelling \citep{Wieringa1992}. Because of this reason redundant calibration is undergoing a renaissance, resulting in further studies by \citet{Noorishad2012,Liu2010,Ali2015,Dillon2016}, showing the applicability and some limitations of redundant calibration in low frequency radio telescopes. More recently, redundant calibration was compared to sky model based calibration by \citet{Li2018}. Despite the inability of redundant calibration to solve for direction dependent effects, it still remains an interesting alternative to calibrate a radio telescope to first order, where sky based calibration can resolve higher order effects.

In this paper we will study the theoretical performance of redundant calibration. We specifically look at how redundant calibration depends on the flux distribution of the sky and positional errors of the antennas. This allows us to determine which regions of the sky should be calibrated with sky-based calibration or redundant calibration, to yield the most accurate and precise result for a given antenna position precision of the array.
We do this by running simulations of redundant calibration in which we calibrate a redundant array with ideal antenna responses on a realistic multi-source sky, while changing the flux and position of a single calibrator source. We compare the distribution of solutions we obtain from these simulations with a theoretical estimate of the sky model calibration bias and an uncertainty due to an incomplete calibration model.

The structure of the paper is as follows: Section \ref{section_incomplete_sky_model} discusses sky model calibration and our analytic description of the impact of an incomplete sky model on the bias and uncertainty of the calibration solutions. Section \ref{red_cal_review} reviews redundant calibration, and describes the set up of the redundant calibration simulations. We discuss the influence of the sky flux distribution on redundant calibration solutions and the impact of position offsets, using a simple 5-element interferometer to demonstrate the fundamental issues of redundant calibration. We conclude our results with a comparison between the bias and uncertainty of redundant calibration, and sky model based calibration for the redundant MWA tiles in Section \ref{Sky_Redundant_Comparison}, and we discuss the implications for the MWA in Section \ref{discussion}.

\section{Sky Model Calibration}
\label{section_incomplete_sky_model}
In this section we describe and derive the impact of an incomplete sky model on the calibration solutions in a sky model based approach. Earlier works studied the effect of calibration on incomplete sky models via analysis and simulations. \citet{Salvini2014} discuss the statistical performance, \citet{Barry2016} study the impact on EoR power spectrum estimation, and \citet{Grobler2014,Wijnholds2016} study its impact on the deconvolution of a 2-point source sky. In this work we compare redundant calibration with theoretical estimates for the bias and uncertainty introduced by calibration on an incomplete sky model.

We can write the measured correlation for a pair of antennas $i$ and $j$ in the absence of noise $C_{ij}$ as a product of the antenna gain factors $g_{i}^{*}$ and $g_{j}$, and the true visibility $V_{ij}$
\begin{equation}
\begin{aligned}
C_{ij} = g_i^{*} g_j V_{ij},
\label{measurement_equation}
\end{aligned}
\end{equation}
the superscript '$^{*}$' indicates complex conjugation. In sky model based calibration we solve for the gains $\mathbf{g}$ by minimizing the difference between our modelled visibilities $\mathbf{M}$ and the measured correlations $\mathbf{C}$.
\begin{equation}
\min_{\mathbf{g}}\| \mathbf{C} - \overline{\mathbf{g}}\mathbf{g}\times \mathbf{M}\|
\label{sky_model_minimization}
\end{equation}
Here, we write the minimization in the most general form, without explicitly choosing a matrix or vector notation for discussion purposes, as we will switch between those later on. The caveat of this approach is that the signals from unmodelled sources are absorbed into the calibration solutions. To understand how this impacts the solutions, we first derive the uncertainty of sky based calibration solutions due to a stochastic sky of point sources and thermal noise. We then use this result to derive the bias due to model incompleteness.  

\subsection{Model Incompleteness Uncertainty}
\label{subsection_incomplete_sky_model_uncertainty}
To derive the minimum uncertainty on the estimated complex gain solutions $g$ we use the Cram\'{e}r-Rao Lower Bound (CRLB) on the estimated gain parameters. Throughout this derivation we assume the model used for calibration is a single point source with flux density $S(\nu)$ located at some location $\mathbf{l}$ in the sky. The model visibility for a given baseline $\mathbf{u}$ at frequency $\nu$ is then given by:
\begin{equation}
\begin{aligned}
M(\mathbf{u}, \nu) = S(\nu)A(\mathbf{l}) \exp [ - 2 \pi i \mathbf{u}\cdot \mathbf{l}], 
\end{aligned}
\label{point_sources_model}
\end{equation}

$A(\mathbf{l})$ is the antenna beam response, which we choose to be a Gaussian. We choose an unmodelled source background described by a broken power-law source count distribution $\mathrm{d}N/ \mathrm{d}S$ \citep{Gervasi2008,Intema2011,Franzen2016,Williams2016} 

\begin{equation}
\begin{aligned}
\frac{\mathrm{d}N}{\mathrm{d}S} =   
\begin{cases}
k_{1} S^{-\gamma_{1}} & \text{if $ S_{\mathrm{low}}\leq S< S_{\mathrm{mid}}$} \\
k_{2} S^{-\gamma_{2}} & \text{if $ S_{\mathrm{mid}}\leq S< S_{\mathrm{high}}$} \\
\end{cases}
,
\end{aligned}
\label{source_counts}
\end{equation}

where $\mathrm{d}N/\mathrm{d}S$ gives the number of sources per $\mathrm{Jy}$ per steradian, and $S$ is the source flux in $\mathrm{Jy}$.
Throughout this paper we will use $k_1 = k_2 = 4100$, $\gamma_1 = 1.59$, $\gamma_2 = 2.5 $, $S_{\mathrm{low}} = 400\, \mathrm{mJy}$, $S_{\mathrm{mid}} = 1\, \mathrm{Jy}$, and $S_{\mathrm{high}} = 5\, \mathrm{Jy}$.
To derive the CRLB on the estimated gain parameters we first compute the Fisher Information Matrix (FIM) $\mathcal{I}$ \citep{Kay1993}. This takes the following form for a complex multivariate normal distribution with mean $M$ and gain independent data covariance $\Sigma_{\mathrm{data}}$; 
\begin{equation}
\begin{aligned}
\mathcal{I}_{i,j} = 2 \mathrm{Re} \Bigg (\frac{\partial \mathbf{M}^H}{\partial g_{i}} \mathbf{\Sigma}_{\mathrm{data}}^{-1} \frac{\partial \mathbf{M}}{\partial g_{j}}\Bigg),
\end{aligned}
\end{equation}
where the superscript `$^H$' denotes the Hermitian transpose, the superscript `$^{-1}$' denotes the matrix inverse, $ \mathbf{M}$ is a vector where each entry is the model visibility of a baseline pair $ij$ for a single frequency channel, and $\mathbf{\Sigma}_{\mathrm{data}}$ is the data covariance matrix. The covariance of the data is the sum of thermal noise variance $\Sigma_{\mathrm{thermal}}$ and the variance of our stochastic background sky $\Sigma_{\mathrm{sky}}$, as we assume the thermal noise is baseline independent and we ignore the compact Fourier beam kernel that creates correlations between closely-spaced baselines.
We describe the thermal noise as
\begin{equation}
\begin{aligned}
\Sigma_{\mathrm{thermal}}= \sqrt{\frac{SEFD}{Bt}}
\end{aligned}
,
\label{noise_variance}
\end{equation}
where $B$ is the bandwidth of a single frequency channel, and $t$ is the integration time of the observation. Throughout this paper we adopt the MWA EoR parameters unless stated otherwise,  $SEFD =10^4\, \mathrm{Jy}$, $B=40\,\mathrm{kHz}$, $t=120\,\mathrm{s}$. For these parameters the thermal noise is $\Sigma_{\mathrm{thermal}} \sim 9 \, \mathrm{Jy}$. We take the expression for the visibility variance for a baseline in a single frequency channel due to a stochastic sky $\Sigma_{\mathrm{sky}}$, from \citet{Trott2016,Murray2017}
\begin{equation}
\begin{aligned}
\Sigma_{\mathrm{sky}} = 2\pi  \frac{\sigma^2}{2} \Bigg(\frac{k_1 }{3 - \gamma _1}\big [S_{\mathrm{mid}}^{3 -\gamma_1} -  S_{\mathrm{low}}^{3-\gamma_1}\big] + \\ \frac{k_2 }{3 - \gamma_2 }\big[S_{\mathrm{high}}^{3 -\gamma_2} -  S_{\mathrm{mid}}^{3-\gamma_2}\big] \Bigg)
\end{aligned}
,
\end{equation}
wherein we assume a flat spectral index of our sources within a single frequency channel, and $\sigma$ is the frequency dependent beam width. Throughout this paper we assume a beam width of  $\sigma = 30^\circ$, similar to the MWA beam at 150 MHz, resulting in a sky variance of $\Sigma_{\mathrm{sky}} \sim 2.5\cdot10^3 \, \mathrm{Jy}$.
Because the noise variance $\Sigma_{\mathrm{thermal}}$ and sky variance $\Sigma_{\mathrm{sky}}$ are baseline independent, the total data covariance matrix $\Sigma_{\mathrm{data}}$ is a diagonal matrix. We can therefore rewrite the FIM elements as;
\begin{equation}
\begin{aligned}
\mathcal{I}_{i,j} = 2 \mathrm{Re} \Bigg( \sum_{\mathrm{n}} \frac{1}{\Sigma_{\mathrm{data}}} \frac{\partial \mathbf{M}^{*}_{n}}{\partial g_i} \frac{\partial \mathbf{M}_{n}}{\partial g_j} \Bigg)
\end{aligned}
,
\end{equation}
where we sum over the data index n. For the CRLB we are only interested in the variance on a gain parameter $\Sigma_{g}$, i.e. we only compute $\mathcal{I}_{m,n}^{-1}$ for $m=n$, which reduces to 
\begin{equation}
\begin{aligned}
\Sigma_{g} = \frac{\Sigma_{thermal} + \Sigma_{\mathrm{sky}}}{2[S(\nu)A(\mathbf{l})]^2(N-1)}
\end{aligned}
,
\end{equation}
where $N-1$ is the number of baselines formed by an antenna in the array.
We note that the variance scales inversely with the number of antennas in the array, and beam-weighted apparent flux density of the modelled source squared. We will use this expression to compare the uncertainty of redundant calibration with sky model based calibration. 

\subsection{Model Incompleteness Bias}
\label{subsection_incomplete_sky_model_bias}
\begin{figure}
\centering
\includegraphics[width=0.19\textwidth]{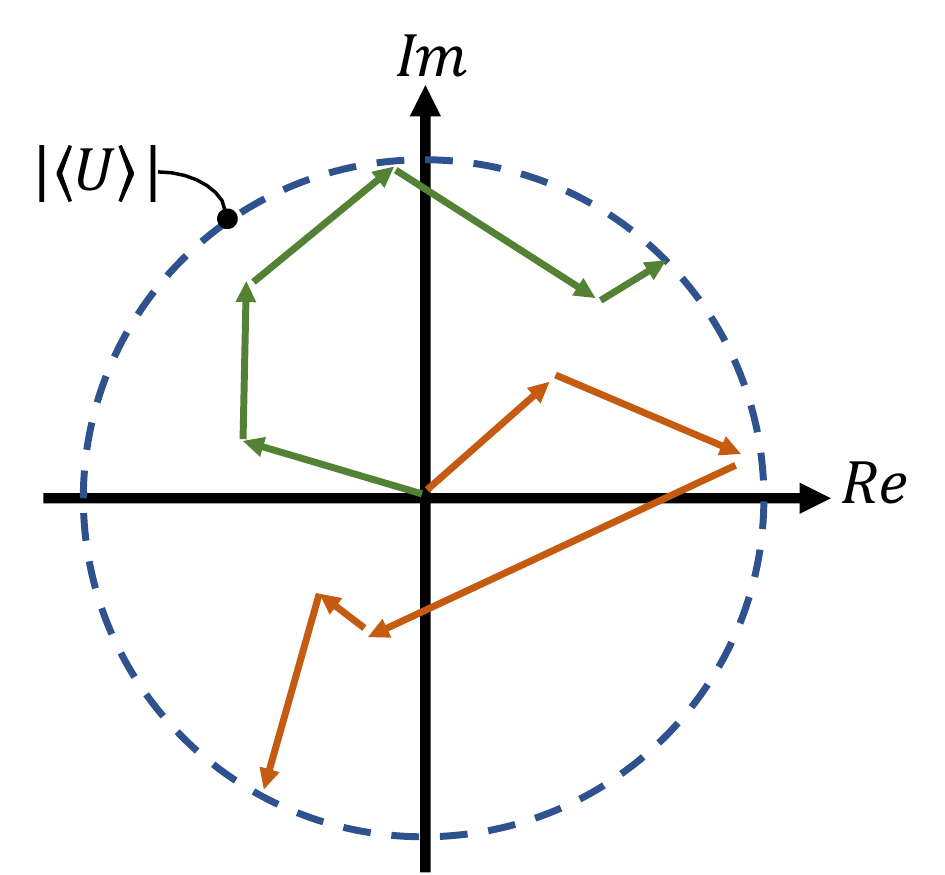}
\includegraphics[width=0.19\textwidth]{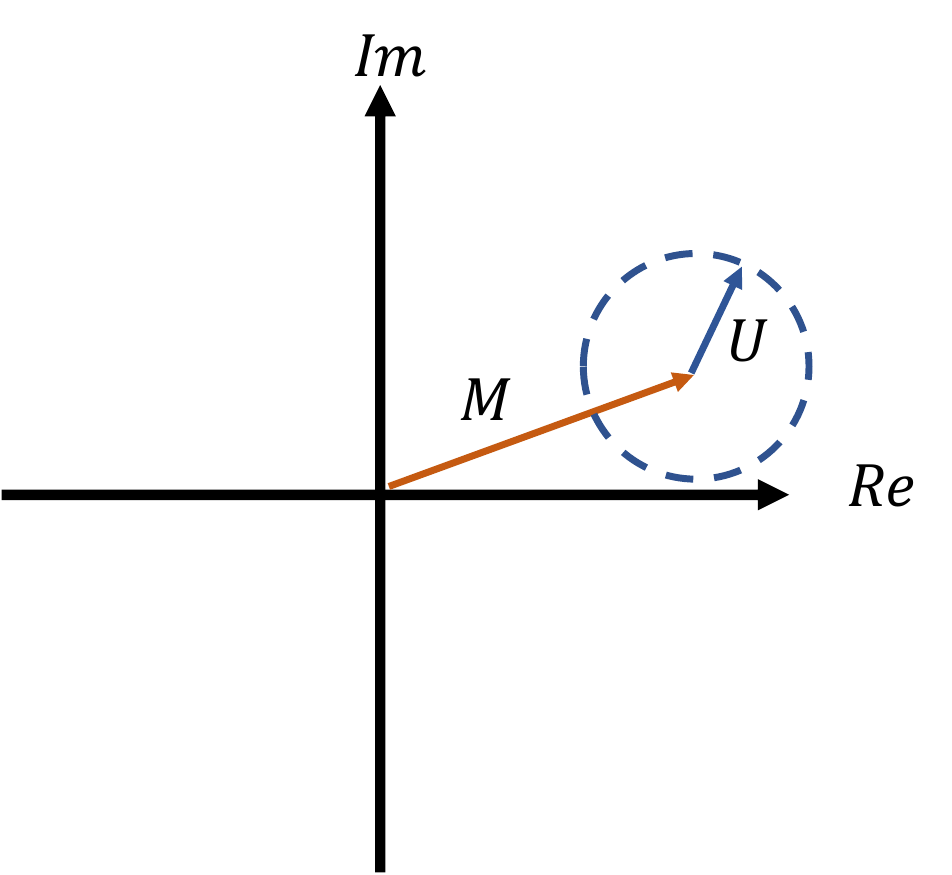}
\caption{\textit{On the left}: an illustration of the random walk through the complex plane of a stochastic sky. \textit{On the right}: The orientation between the modelled visibility $\mathbf{M}$ and the unmodelled visibility contribution $\mathbf{U}$.}
\label{Unmodelled_Modelled_sky_illustration}
\end{figure}

To derive an expression for the bias, i.e. the mean deviation from the true solutions introduced by the model incompleteness, we follow \citet{Wijnholds2016} and reformulate Equation \ref{sky_model_minimization} explicitly in terms of visibility matrices and gain vectors;
\begin{equation}
\min_{\mathbf{g}}\| \mathbf{C} - \mathbf{g}\mathbf{M}\mathbf{g}^H\|.
\label{sky_model_matrix}
\end{equation}
$\mathbf{C}$ and $\mathbf{M}$ are matrices containing the measured and modelled visibilities, e.g. $\mathbf{C}_{ij}$ is the measured visibility between antenna $i$ and $j$, and the vector $\mathbf{g}$ contains the complex antenna gains. We ignore the auto-correlations, therefore, the diagonals of $\mathbf{M}$ and $\mathbf{C}$ are zero, and if we ignore the noise we can write the measurements $\mathbf{C}$ in terms of the modelled $\mathbf{M}$ and unmodelled $\mathbf{U}$ sky visibilities. We can also write the gain vector $\mathbf{g}$ as a sum of the true gains $\mathbf{g}_t$ and a deviation introduced by the calibration process $\Delta \mathbf{g}$.

\begin{equation}
\begin{aligned}
\mathbf{C} &= \mathbf{g}_t (\mathbf{M} + \mathbf{U})\mathbf{g}_t^{H} \\
\mathbf{g} &= \mathbf{g}_t + \Delta \mathbf{g}
\end{aligned}
\label{calibration_error}
\end{equation}
Furthermore we can use the \textit{Hadamard product} $\odot$, i.e. the element-wise product, to rewrite Equation \ref{sky_model_matrix} into
\begin{equation}
\min_{\Delta \mathbf{g}}\| \mathbf{g}_t \mathbf{g}_t^H \odot \mathbf{U} - (\mathbf{g}_t \Delta \mathbf{g}^H + \Delta \mathbf{g}\mathbf{g}_t^H)\odot \mathbf{M}\|
\label{incomplete_sky_model_simplified}
\end{equation}
where we have dropped all higher order terms of $\Delta \mathbf{g}$. \citet{Wijnholds2016} derive an approximate closed form solution for $\Delta \mathbf{g}$ by rewriting Equation \ref{incomplete_sky_model_simplified} into a least squares form. We will take the solution as the conclusion of this short review, and point the interested reader to their work for the detailed derivation. The closed form solution takes the following form
\begin{equation}
\begin{aligned}
\begin{bmatrix}
\Delta \mathbf{g} \\
 \Delta\mathbf{g}^{*}
\end{bmatrix}
 \approx 
\begin{bmatrix}
 \mathbf{A} & \mathbf{B} \\
 \mathbf{C} & \mathbf{D}
 \end{bmatrix}^{-1}
 \begin{bmatrix}
 \mathbf{E}\mathbf{g}_t \\
 \mathbf{F}\mathbf{g}_t
 \end{bmatrix}
\end{aligned}
.
\label{wijnholds_solution}
\end{equation}
The block matrices are given by
\begin{equation}
\begin{aligned}
\mathbf{A}&= \mathbf{M}^{*} \mathbf{G}_t \mathbf{G}_t^H \mathbf{M}^{*} \odot \mathbf{I} & \mathbf{B}&= \mathbf{M}^{*} \mathbf{G}_t \odot \mathbf{G}_t \mathbf{M} \\
\mathbf{C}&= \mathbf{B}^{*} & \mathbf{D}&= \mathbf{A}^{*}\\
\mathbf{E}&= \mathbf{M}^{*}\mathbf{G}_t \mathbf{G}_t ^H \mathbf{U}^{*} \odot \mathbf{I} & \mathbf{F}&= \mathbf{G}_t^H \mathbf{U}^{*} \odot \mathbf{M} \mathbf{G}_t ^H \\
\end{aligned}
,
\label{wijnhold_appendum}
\end{equation}
where $\mathbf{G}_t = \mathrm{diag}(\mathbf{g})$, and $\mathbf{I}$ is the identity matrix.
Here, $\mathbf{A}$ and $\mathbf{B}$ encode the total modelled power summed over baselines, and the power in an individual baseline, respectively, whereas $\mathbf{E}$ and $\mathbf{F}$ are the equivalent expressions for the unmodelled power. Intuitively, these matrices describe the additional bias in the solutions from correlations between the model and the residual signal, and the overall power ratio of model to unmodelled sky. Minimising both of these bias terms is desirable for good sky-based calibration.
We can use Equation \ref{wijnholds_solution} to derive the mean gain offset $\langle \Delta \mathbf{g} \rangle$ in the case that our sky model consists of a single point source in the presence of a more complicated sky.
\begin{equation}
\begin{aligned}
\Bigg \langle
\begin{bmatrix}
\Delta \mathbf{g} \\
 \Delta\mathbf{g}^{*}
\end{bmatrix}
\Bigg \rangle
 \approx 
\begin{bmatrix}
 \mathbf{A} & \mathbf{B} \\
 \mathbf{C} & \mathbf{D}
 \end{bmatrix}^{-1}
 \begin{bmatrix}
 \langle \mathbf{E} \rangle\mathbf{g}_t \\
\langle \mathbf{F} \rangle \mathbf{g}_t
 \end{bmatrix}
\end{aligned}
,
\label{averaged_wijnholds_solution}
\end{equation}
with 
\begin{equation}
\begin{aligned}
\langle \mathbf{E} \rangle &= \mathbf{M}^{*}\mathbf{G}_t \mathbf{G}_t ^H   \langle \mathbf{U}\rangle^{*}  \odot \mathbf{I} \\
\langle \mathbf{F} \rangle &= \mathbf{G}_t^H \langle \mathbf{U}\rangle^{*} \odot \mathbf{M} \mathbf{G}_t ^H \\
\end{aligned}
.
\label{wijnholds_appendum_averaged}
\end{equation}
We can parametrize the mean unmodelled visibility contribution of our stochastic sky $\langle \mathbf{V}_u \rangle$ using the sky visibility variance $\Sigma_{\mathrm{sky}}$. If we consider the Fourier transform of each point source as a phasor in the complex plane $(\mathrm{Re}, \mathrm{Im})$, we can consider a stochastic sky of point sources as a random walk through this plane (see Figure \ref{Unmodelled_Modelled_sky_illustration}). Each point source contributes a new complex phasor to our total unmodelled sky phasor. The path length of this random walk, i.e. the total amplitude of our unmodelled visibility, is on average given by the root mean square of distribution from which the phasors are drawn. In our analysis we assume this to be a Gaussian distribution, therefore, the unmodelled visibility amplitude equates to the variance. 

Now, we have yet to explore (are still left with) the orientation of the unmodelled visibility component, since every net orientation has equal probability.  For our calculations we choose the unmodelled visibility to have an angle of $45^{\circ}$, with respect to the model visibility as a measure for some average offset introduced into our visibility amplitude and phase angle. Even though the true phase angle of the unmodelled visibility is uniformly distributed, we find that this approximation yields comparable results to Monte Carlo simulations with a distribution of phases. We will use these results in Section \ref{Sky_Redundant_Comparison} where we compare the results from our redundant calibration simulations with the theoretical performance of sky model based calibration.

\section{Redundant Baseline Calibration}
\label{red_cal_review}
In this section we provide a short review of \citet{Wieringa1992} and \citet{Liu2010} to highlight the key features of redundant calibration. Considering the simplest redundant case, which is a 5 element linear array with equal spacings $\Delta \mathrm{x}$ between the antennas, we have four unique baselines. Four baselines at the shortest spacing, three at 2$\Delta \mathrm{x}$, two at 3$\Delta \mathrm{x}$ and only one at 4$\Delta \mathrm{x}$. Using only the first two sets of baselines, we can create a determined system of linear equations, solving for the 2 unknown visibilities for each set of redundant baselines and 5 unknown antenna gains. One way of doing this is the \texttt{logcal} algorithm, in which we take the logarithm of Equation \ref{measurement_equation}, while noting that each quantity is a complex number $\lvert g \rvert \exp[i\phi]$ with amplitude $\lvert g \rvert$ and phase $\phi$. This yields two equations,
\begin{equation}
\begin{aligned}
\mathrm{ln} \lvert C_{ij} \rvert &=  \mathrm{ln} \lvert g_{i} \rvert + \mathrm{ln} \lvert g_{j} \rvert  + \mathrm{ln} \lvert V_{ij} \rvert \\
\mathrm{arg} \lvert C_{ij} \rvert &= \phi_{j} - \phi_{i}  + \mathrm{arg} \lvert V_{ij} \rvert\\
\end{aligned}
\label{logcal}
\end{equation}
where $\lvert g_{i} \rvert$  is the gain amplitude and $\phi_{i}$ is the gain phase of antenna $i$, $\mathrm{ln} \lvert V_{ij} \rvert$ is the true visibility amplitude and $\mathrm{arg} \lvert V_{ij} \rvert$ is the true visibility phase measured by a baseline pair $i$ and $j$. Because the amplitude and phase decouple, we can rewrite this into two different matrix equations that can be solved independently, see Equation \ref{matrix_equation}.

\begin{equation}
\begin{aligned}
\mathbf{c}_{\alpha} =\mathbf{A}_{\alpha}\mathbf{x}_{\alpha}\\ 
\end{aligned}
,
\label{matrix_equation}
\end{equation}
where the index $\alpha = \{\eta,\phi\}$, $\eta$ for the amplitude equations and $\phi$ for the phase equations, $\mathbf{c}$ are the measured correlation amplitude and phase vectors, $\mathbf{x}_{\eta}$ contains the gain amplitude $\mathrm{ln} \lvert g_{i} \rvert$ and visibility amplitude $\mathrm{ln} \lvert V_{ij} \rvert$, $\mathbf{x}_{\phi}$ contains the gain phase $\phi_{j}$ and visibility phase $\mathrm{arg} \lvert V_{ij} \rvert$.
$\mathbf{A}$ is the matrix that maps the gain and visibility into the measured correlations. Equation \ref{phase_matrix_equation} shows this explicitly for the phase,
\begin{equation}
\begin{aligned}
\begin{pmatrix}
\mathrm{arg} \lvert c_{12} \rvert \\
\mathrm{arg} \lvert c_{23} \rvert \\
\mathrm{arg} \lvert c_{34} \rvert \\
\mathrm{arg} \lvert c_{45} \rvert \\
\mathrm{arg} \lvert c_{13} \rvert \\
\mathrm{arg} \lvert c_{24} \rvert \\
\mathrm{arg} \lvert c_{35} \rvert \\
\end{pmatrix} 
=
\begin{pmatrix}
-1 & 1 & 0 & 0 & 0 & 1 & 0 \\
0 & -1 & 1 & 0 & 0 & 1 & 0 \\
0 & 0 & -1 & 1 & 0 & 1 & 0 \\
0 & 0 & 0 & -1 & 1 & 1 & 0 \\
-1 & 0 & 1 & 0 & 0 & 0 & 1 \\
0 & -1 & 0 & 1 & 0 & 0 & 1 \\
0 & 0 & -1 & 0 & 1 & 0 & 1 \\
\end{pmatrix}
\begin{pmatrix}
\phi_{1} \\
\phi_{2} \\
\phi_{3} \\
\phi_{4} \\
\phi_{5} \\
\mathrm{arg} \lvert v_{1} \rvert \\
\mathrm{arg} \lvert v_{2} \rvert \\
\end{pmatrix}
\end{aligned}
\label{phase_matrix_equation}
\end{equation}
$\mathrm{arg} \lvert v_{1} \rvert $ is the phase of the visibility measured by the $\Delta x$ spacings and for the $2 \Delta x$ spacings we have $\mathrm{arg} \lvert v_{2} \rvert$.
This system is, however, degenerate and needs to be constrained by setting a reference antenna for which the amplitude gain and phase gain are specified. In the specific case of phase calibration we need two additional constraints because a tilt in the array is equivalent to a rotation of the sky.
\begin{equation}
\begin{aligned}
0 = \mathrm{ln} \lvert g_{1} \rvert \quad	& 0 = \phi_1  \\
					& 0 = \Sigma x_i \, \phi_i  \\
					& 0 = \Sigma y_i \, \phi_i 
\end{aligned}
,
\label{redundant_calibration_degeneracy}
\end{equation}
where $x_i$ and $y_i$ represent the ideal redundant position coordinates of the antenna within the array. Now that the degeneracies have been broken both system of equations can be solved using the general least square solution for a linear equation:
 $\mathbf{\hat{x}} = [\mathbf{A}^{T}\mathbf{A}]^{-1}\mathbf{A}^{T}\mathbf{c}$. 
 
Another way of linearising Equation \ref{measurement_equation} is \texttt{lincal}, in which we take a Taylor expansion around solution guesses $g_i^0$ and $v_{ij}^0$ of the true solutions $g_i$ and $v_{ij}$. This yields one single equation in which we solve for the gains in their complex forms,
\begin{equation}
\begin{aligned}
c_{ij}  &= g_i^0 g_j^{0*}v_{ij}^0 + g_j^{0*}v_{ij}^0\Delta g_{i} + g_i^{0}v_{ij}^0\Delta g_{j}+g_{i}^0 g_j^{0*}\Delta v_{ij}
\end{aligned}
\label{lincal}
\end{equation}
In \texttt{lincal} we solve for the differences between true solutions and the guesses $\Delta g_{i}$ and $\Delta v_{ij}$. Allowing us to iteratively correct our guesses. Similarly to \texttt{logcal} we can rewrite this into a matrix equation containing the real and imaginary components of the gains and visibilities, for the details see the appendix of \citet{Li2018}. Current implementations of redundant calibration use \texttt{logcal} to find an initial estimate and further refine the solutions with \texttt{lincal}. \citep{Zheng2014} In this work we will do the same. 

\subsection{Simulating the Bias and Uncertainty}
\label{simulation_description}
To estimate the bias and uncertainty of redundant calibration we simulate the calibration of the antennas in a (nearly) redundant array. In our simulations we define a group of redundant baselines when they lie within 1/6$\lambda$ of each other in the $uv$-plane. This is well within the linear regime of sinusoidal centred at $\exp{[2\pi ul]}$. However, we will show that deviations from non-redundancy within this threshold impact the calibration accuracy and precision. We also assume a Gaussian beam, similar to our sky model derivation, and assume the beams are identical for each antenna. This is not strictly true for phased arrays \citep{Wijnholds2010}, however, it suffices as a first-order approximation.

We generate a background of radio sources with a flux distribution according to Equation \ref{source_counts}, and uniformly distribute them over the sky. Finally, we add a source with arbitrary flux and location, similar to the calibrator source in Section \ref{section_incomplete_sky_model}. These sources are gridded onto an $(l,m)$-grid, and Fourier transformed to generate visibilities using \texttt{powerbox} \citep{Murray2018}, a tool written to simulate EoR datasets and forward-model them to power spectra. We interpolate the visibilities to produce the measurements for each baseline. Finally, we assume Gaussian-distributed noise in the real and imaginary components with a variance according to Equation \ref{noise_variance}. These visibilities are then passed to our redundant calibration module, which is a direct implementation of the algorithm described in Section \ref{red_cal_review}. The code is publicly available \citep{Joseph2018a}.

\begin{figure*}[!htp]
  \centering
\includegraphics[width=1\textwidth]{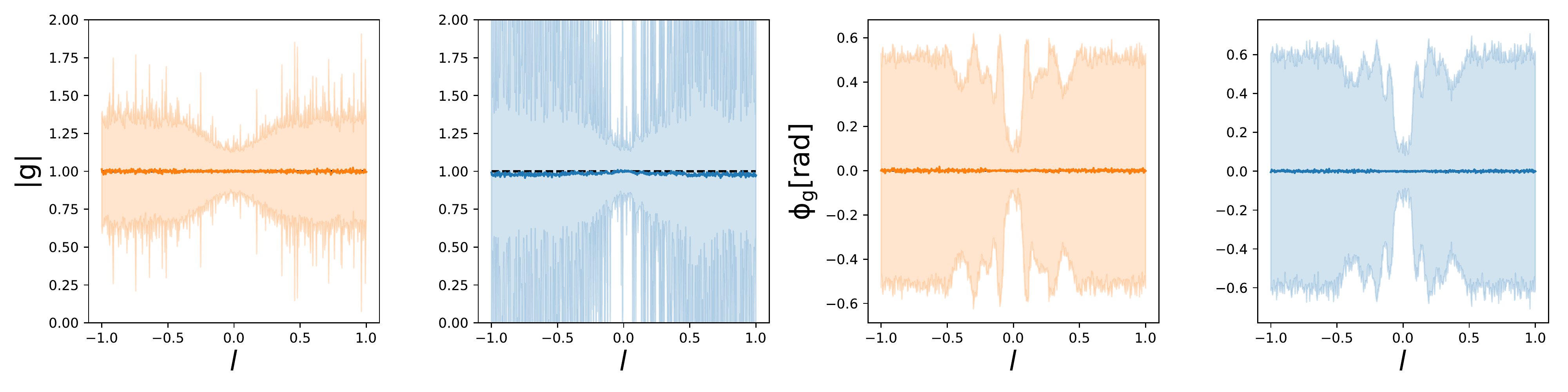}
\caption {From left to right: the \texttt{logcal} amplitude gain, the \texttt{logcal +lincal} amplitude gain, the \texttt{logcal} phase gain solutions, and \texttt{logcal +lincal} phase gain solutions for a single antenna in an nearly redundant 5 element interferometer as a function of strong source position $l$. The dark line represents the mean of the solutions, the shaded area indicates the 1-sigma solutions variance. The amplitude solution variance inversely follows the shape of the beam, i.e. the variance increases when the beam response decreases. The mean of the phase solutions fluctuates around $\phi =0$ and jumps along with the variance at so-called phase wrapping points, which are further explained in the text.}
\label{5_element_StaRCal}  
\end{figure*}

\subsection{The Sky Dependent Uncertainty}
\label{sky_dependence}
We first study the influence of the sky, and show how it affects the uncertainty of the estimated gain solutions. We start out with a simple sky model of a statistical background sky while moving a high flux density source with respect to phase centre and trying to calibrate on each realization of the sky. This allows us to study the performance of redundant calibration in drift-scan mode
, and simultaneously study the performance of redundant calibration in the MWA EoR fields, which depending on the field have strong in-beam sources. Figure \ref{5_element_StaRCal} shows the dependence of the calibration solutions for an ideal interferometer, i.e. perfect gains $g = 1$ and perfect redundancy, as a function of source position in terms of the direction cosine $l$, the native interferometry sky coordinate. We show the results for a pure \texttt{logcal}-calibration, similar to \citet{Wieringa1992}, and for a \texttt{logcal + lincal}-calibration.

The results show that overall the solution variance for both the gain amplitude solutions behave better when the strong source is near the centre of the beam, because the signal to noise ratio (S/N) is higher at the pointing centre. We do note that our implementation of the \texttt{lincal} algorithm seems slightly biased in the presence of noise, the mean of the solution is 1\% below the true value. We filtered out $< 1\%$ of the solution realisations due to bad convergence, i.e. solutions with unrealistically high gain amplitudes.

The gain phase solutions show a similar dependence with some additional structure in the variance due to a problem which is inherent to \texttt{logcal}: phase wrapping. The \texttt{logcal} implementation can only determine phases between $- \pi \leq \phi \leq \pi$, in which the $\arctan$ is defined. When a certain redundant set of baselines measures a visibility phase of $\lvert \phi_{v}\rvert = \pi$, due to the location of the dominant source on the sky, the solutions become very sensitive to noise. The visibility phase starts to ``jump'' between $-\pi$ and $\pi$ causing large variances in the phase calibration solutions. These phase wrapping points can be determined by solving $2 \pi u l = \pi n$ for odd numbers of n, i.e. solving for the source coordinate $l$ when a given baseline with length $u$ measures a phase of $\pi$. 

We can understand the effect of phase wrapping by adding a noise vector $\mathbf{n_\alpha}$ to the measurement equation, see equation \ref{matrix_equation_noise}.
\begin{equation}
\begin{aligned}
\mathbf{c}_{\alpha} =\mathbf{A}_{\alpha}\mathbf{x}_{\alpha} + \mathbf{n}_{\alpha}  \\ 
\end{aligned}
\label{matrix_equation_noise}
\end{equation}
The phase noise $\mathbf{n}_\phi \propto N/S$ \citep{Liu2010}, however, when the measured visibility phase approaches $\pi$, this noise vector diverges $\lvert \mathbf{n}_\phi \rvert \rightarrow 2\pi$. Because the noise of a single baseline is mixed into all solutions when estimating $\mathbf{\hat{x}}$ we get large offsets in the calibration solutions:

\begin{equation}
\begin{aligned}
\mathbf{\hat{x}} =\mathbf{x}_{\alpha} + [\mathbf{A}^{T}\mathbf{A}]^{-1}\mathbf{A}^{T} \mathbf{n}_{\alpha}  \\ 
\end{aligned}
.
\label{matrix_noise_mixing}
\end{equation}

\begin{figure}[!htp]
  \centering
\includegraphics[width=0.35\textwidth]{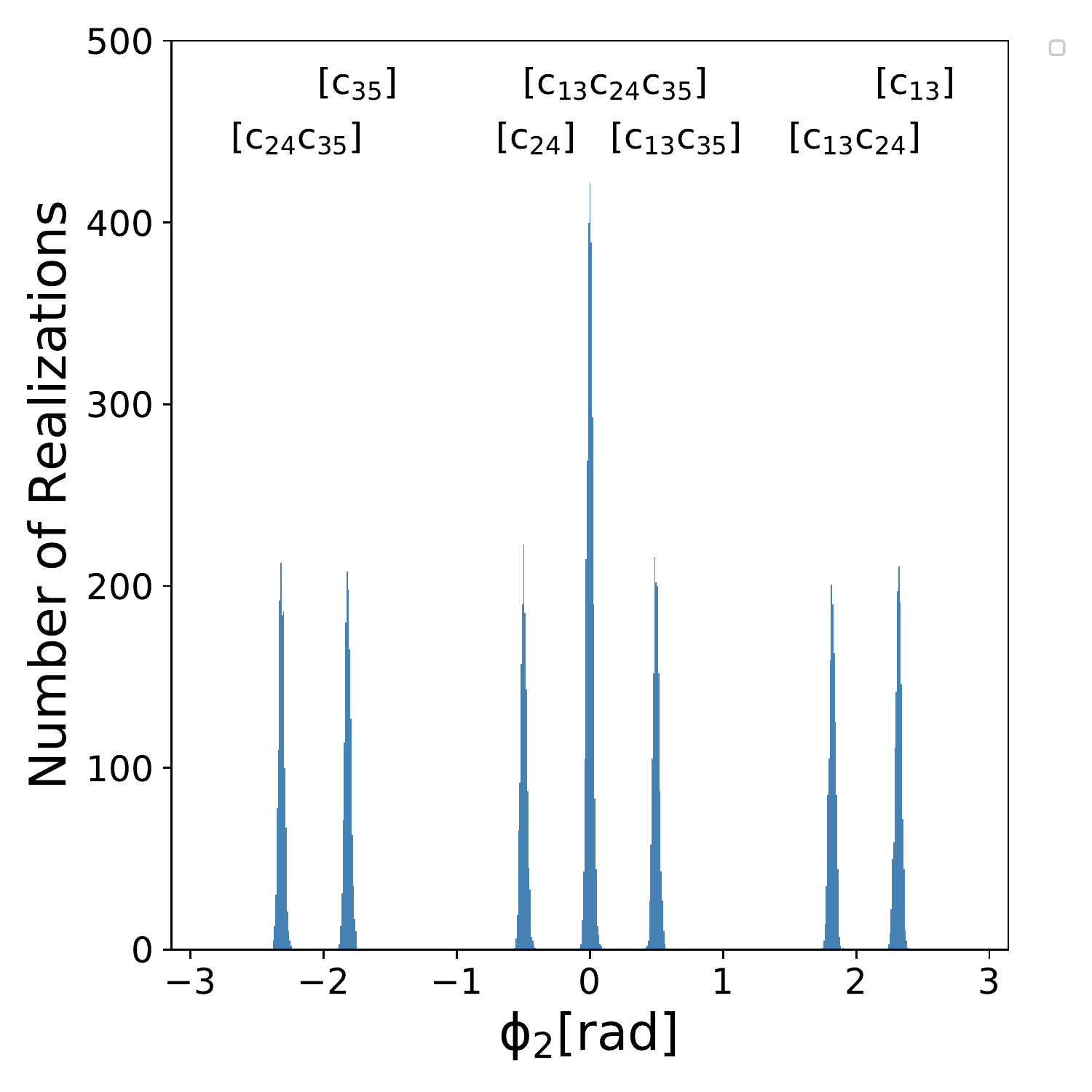}
\caption {The distribution of gain phase solutions for antenna 2 at the first phase wrapping point $l \sim 0.1$. The discrete set of solutions peaks is caused by phase wrapping in the set of three long baselines. Each peak represents a combination of phase wrapping baselines as marked above the peak}
\label{noisy_pdf}  
\end{figure}

Looking at the \texttt{logcal + lincal} solutions in Figure \ref{5_element_StaRCal} we see similar behaviour in the phase solutions and amplitude solutions. We also note that our implementation of the algorithm seems to be very sensitive to noise. This results in large variations in the amplitude solutions when the dominant source moves away from phase center. 
Figure \ref{5_element_StaRCal} shows the mean and the variance of the solutions, even though the solutions do not strictly follow a Gaussian distribution. Figure \ref{noisy_pdf} shows the distribution of \texttt{logcal} solutions for antenna 2 at the first phase wrapping point. The distribution has 7 distinct peaks, each peak represents a combination of phase wrapping baselines. The first phase wrap occurs in the set of three long redundant baselines, and therefore there are at maximum $\Sigma_{k=1}^n n!/(k!(n-k)!)$ different combinations and solutions peaks possible. The actual spectrum depends on the array geometry, which is encapsulated in the matrix $\mathbf{A}$. These solutions are the starting point for the \texttt{lincal} algorithm, and our implementation of it is not able to recover the true solutions when given a bad starting point.

We attempted to circumvent the phase wrapping of a specific baseline in a single channel by extending the \texttt{logcal} algorithm to incorporate the measurements of neighbouring frequency channels while assuming the gain solutions remain the same. However, due to the same mixing that takes place in the single frequency channel implementation, phase wrapping will still remain a problem unless a clever selection of frequencies is used to circumvent phase wrapping. We discuss this in the Appendix. We also note that current implementation do apply a pre-calibration step to unwrap the visibility phases by averaging over baselines within a redundant group \citep{Zheng2014}, or by using the products of visibilities to construct a system equations to solve for the phases \citep{Li2018}.

\subsection{The Position Offset Bias}
\label{near_redunancy}
\begin{figure*}[!ht]
  \centering
\includegraphics[width=1\textwidth]{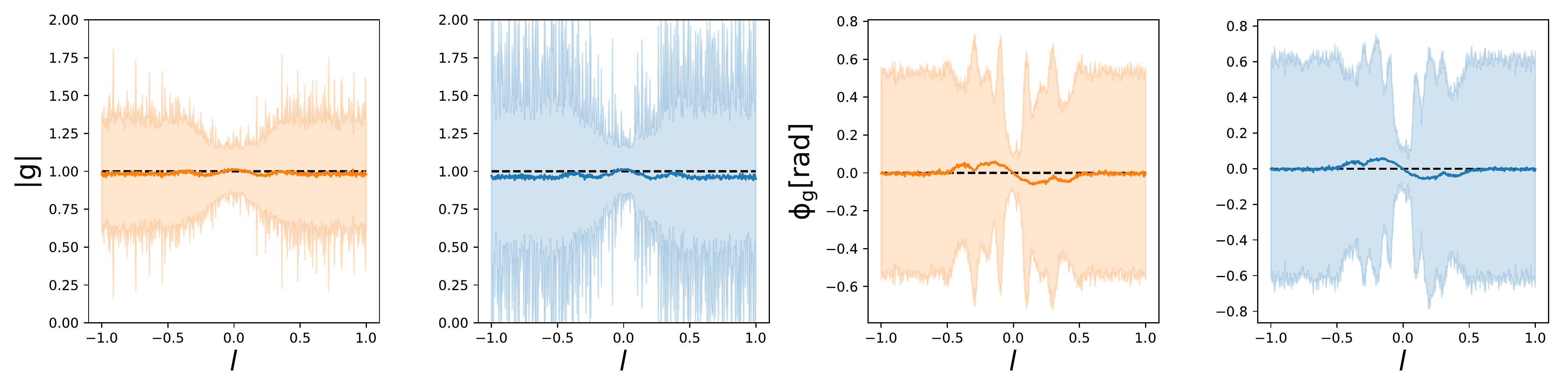}
\caption {From left to right: the \texttt{logcal} amplitude gain and phase gain solutions, the \texttt{logcal +lincal} amplitude gain and phase gain solutions for a single antenna in an nearly redundant 5 element interferometer as a function of strong source position $l$. The dark line represents the mean of the solutions, the shaded area indicates the 1-sigma solutions variance. The amplitude solution variance inversely follows the shape of the beam, i.e. the variance increases when the beam response decreases. The mean of the phase solutions generally fluctuates around $\phi =0$ and deviates along with the variance at so-called phase wrapping points, which are further explained in the text. The additional position offset, which causes a phase offset from the ideal redundant phase, is absorbed into the solutions causing a slope.}
\label{5_element_StaRCal_offset}  
\end{figure*}
In the previous section we described the results for an ideal radio interferometer. However, in reality all antennas will have slight position offsets from their perfectly redundant positions. To understand the impact of positions offset we simulate redundant calibration under the same conditions as before, but now we offsetting one antenna in the x-direction by $\delta x = 20\, \mathrm{cm}$. The results are shown in Figure \ref{5_element_StaRCal_offset}. 

We can clearly see that both the amplitude and phase solutions are affected by the position offset. We can understand the oscillatory behaviour of the amplitude solutions by returning to the complex plane. Imagine the complex visibility of the main calibrator as measured by a baseline as a phasor in this plane, we can think of the total sum of background sources as a similar phasor.  Each redundant baseline should measure the same amplitude of the sum of these phasors. However, due to the non-redundancy introduced by position offsets the non-redundant pairs measure a different amplitude, this difference propagates through to the solutions. As the primary source moves across the sky its phasor will rotate in the complex plane, constructively and destructively interfering with the background visibility creating this oscillatory behaviour. These oscillations are dampened as the primary source becomes attenuated as it moves outside of the primary beam.The behaviour of the  mean phase solutions can be explained in a similar fashion, using the phase of the phasors rather than the amplitude.

For low-N arrays similar to this 5 element toy model this error propagates to all antennas solutions, due to the coupling of all gain solutions to the visibilities. However, when increasing the number of antennas in the array the coupling becomes weaker increases as the number of measurements increases.

\begin{equation}
\begin{aligned}
\mathbf{c}_{\alpha} =\mathbf{A}_{\alpha}\mathbf{x}_{\alpha} + \mathbf{n}_{\alpha} + \mathbf{b}_{\alpha}  \\ 
\end{aligned}
\label{matrix_equation_offset}
\end{equation}

\begin{figure}[htp!]
 \centering
\includegraphics[width=0.35\textwidth]{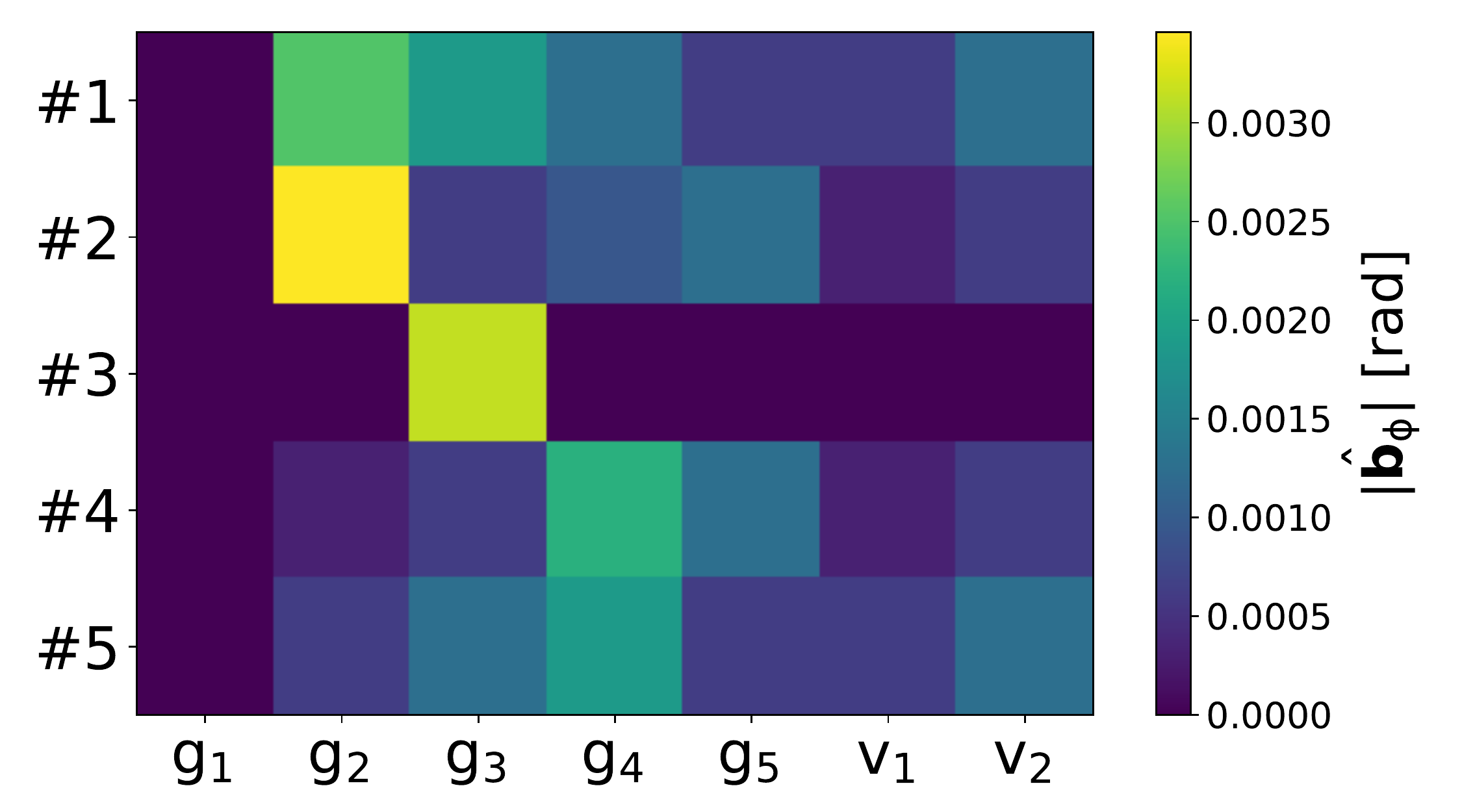}
\caption {The magnitude of the offset residual $\hat{\mathbf{b}}$ for an offset in a given antenna. The first column shows the residual when offsetting the first antenna, i.e. the reference antenna, the second column shows the result for the second antenna etc. We see that offsetting the 3rd antenna, e.g. the middle antenna, impacts only its calibration solution. Its solutions have the strongest constraints due to its baseline participation. It participated in all baselines groups and has a high participation number in each group. }
\label{offset_residual}  
\end{figure}

Inverting this equation using the standard least square solution, and not taking into account this extra term, leaves us with an additional residual. We can calculate these residuals for different antenna offsets. Figure \ref{offset_residual} shows the magnitude of the offset residuals $\hat{\mathbf{b}}$ in the phases of the estimated gain and visibilities when offsetting different antennas by the same amount.  These results show that offsetting the antenna with the highest baseline participation does not propagate to all antenna solutions and leaves the visibilities unaffected. Offsetting the reference antenna, in this case antenna 1, has the strongest impact on the solutions of all other antennas. This implies that the choice of reference antenna is not as arbitrary as one might think.\\

\section{Comparing Sky and Redundancy Based Calibration}
\label{Sky_Redundant_Comparison}
\begin{figure*}[htp]
 \centering
\includegraphics[width=0.48\textwidth]{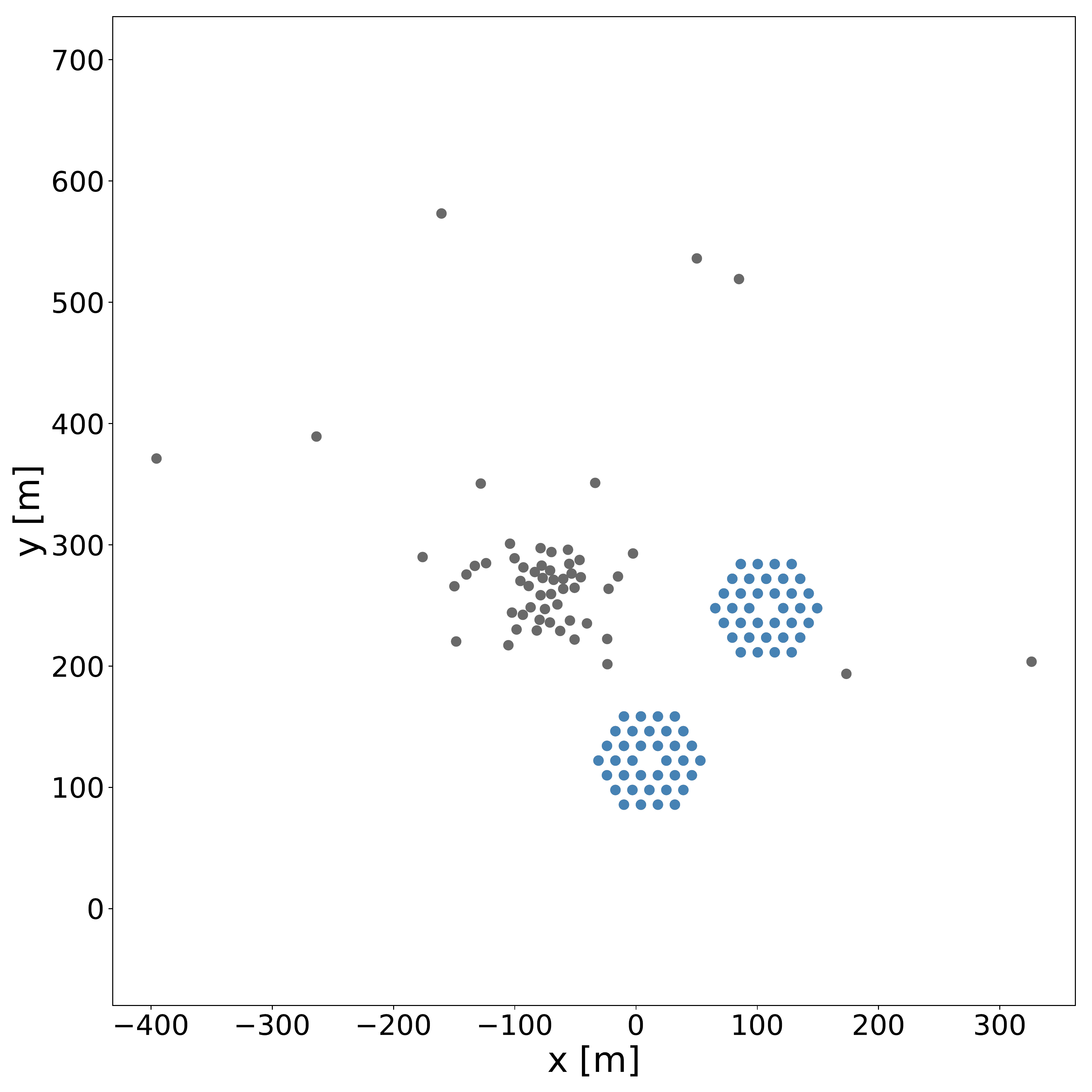}
\includegraphics[width=0.48\textwidth]{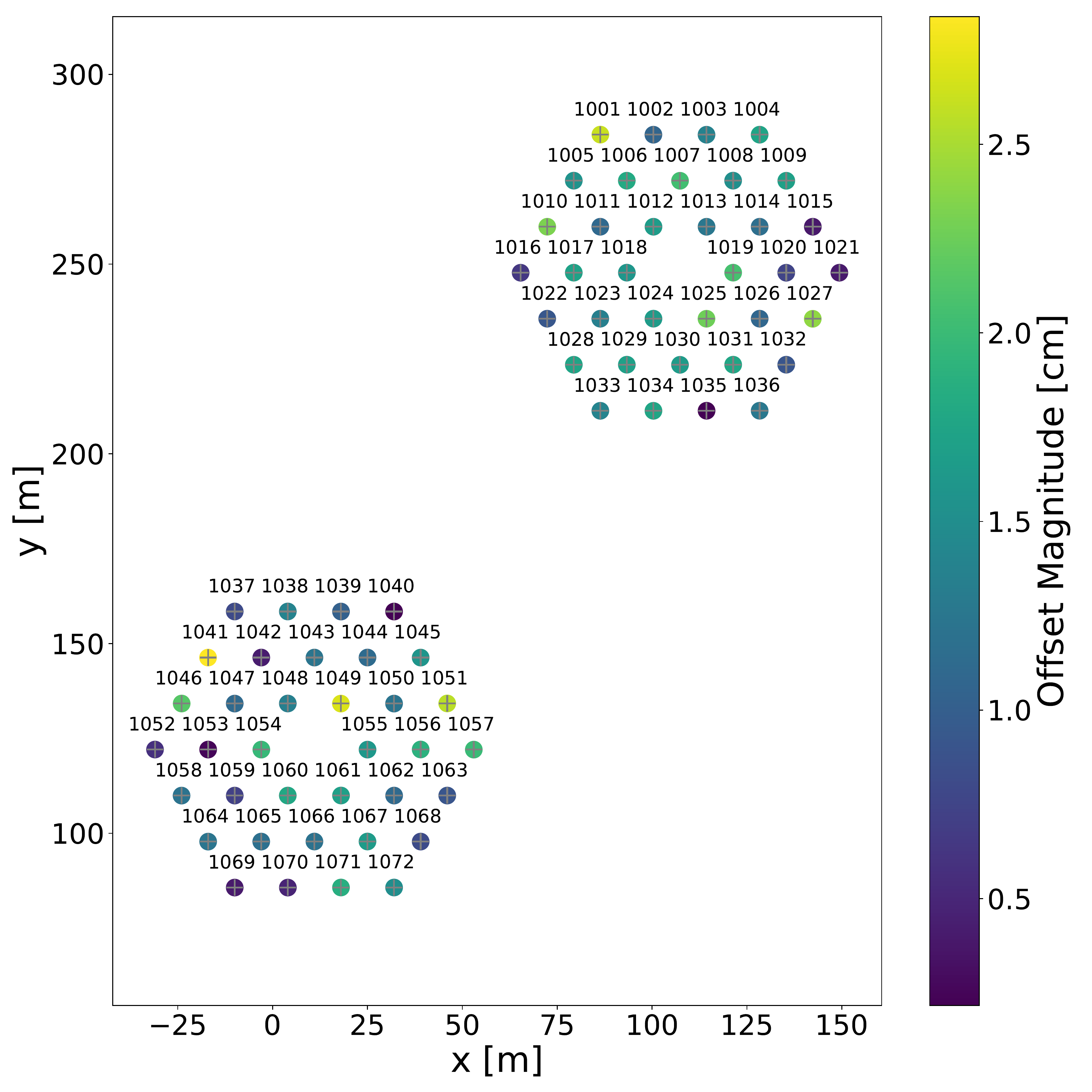}
\caption {\textit{left}: The MWA antenna lay-out, with MWA hex antennas in blue and the random antennas in black. \textit{Right}: a zoom-in plot on the MWA hex tiles. The colour of each tile indicates its offset from its redundant position. We see that the position deviations from redundancy do not exceed 3 cm. }
\label{mwa_hexes}  
\end{figure*}

Having varied several parameters within our redundant calibration simulations we can now move forward and apply this formalism to the MWA "hexes". The hexes contain 72 antenna tiles arranged in 2 hexagons, see Figure \ref{mwa_hexes}. The shortest baseline, defining the hexagonal lattice, has a length of 14 metres. The hexagons are also placed to be redundant with each other, i.e. they have the same orientation. This provides extra sensitivity on scales relevant for the EoR experiment, and adds redundancy for calibration purposes.  Due to a lack of redundant baselines connecting one hex to the reference antenna in the other hex, we either need to invoke another degeneracy parameter to encapsulate a phase offset between the two or calibrate them separately. \citet{Li2018} calibrate the hexes simultaneously, however, for simplicity and speed we calibrate a single hex in our simulations. A single hex forms 630 baselines, of which 601 are redundant, organized in 71 redundant groups.\footnote{Theoretically we can also include non-redundant antennas in the calibration, as long as the number of unknowns is lower than the number of measurements. For each redundant hex in the MWA we can add 6 non-redundant tiles before the system becomes unsolvable.}\\
Figure \ref{mwa_hexes} also shows the non-redundancy of each antenna within the hexes, the antennas are placed with an accuracy on the order of centimetres. This is an order of magnitude below the redundant calibration threshold of $1/6\lambda$ at 150 MHz, i.e. $\sim$30~cm.

\begin{figure*}[htp!]
  \centering
\includegraphics[width=1\textwidth]{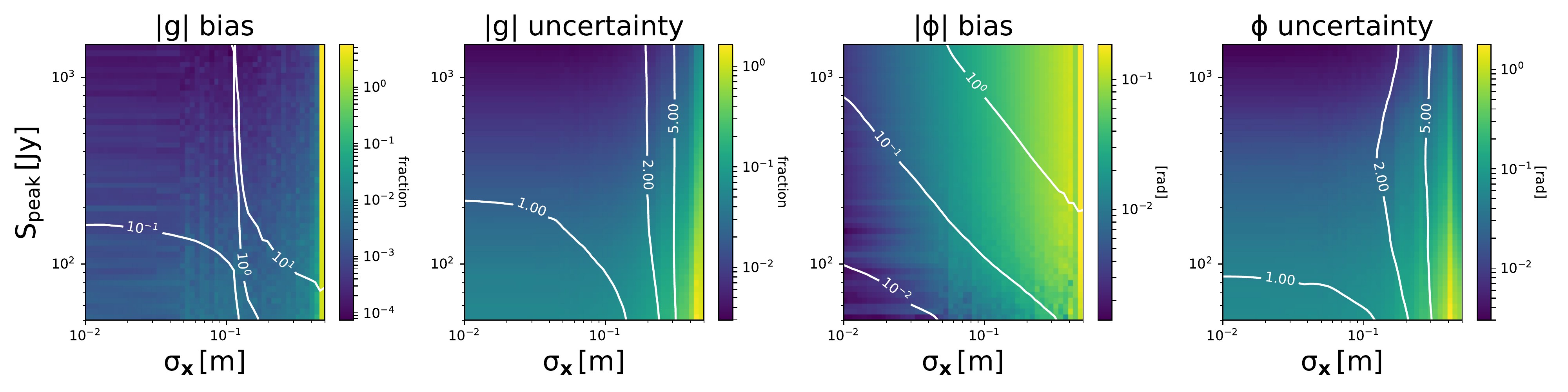}
\includegraphics[width=1\textwidth]{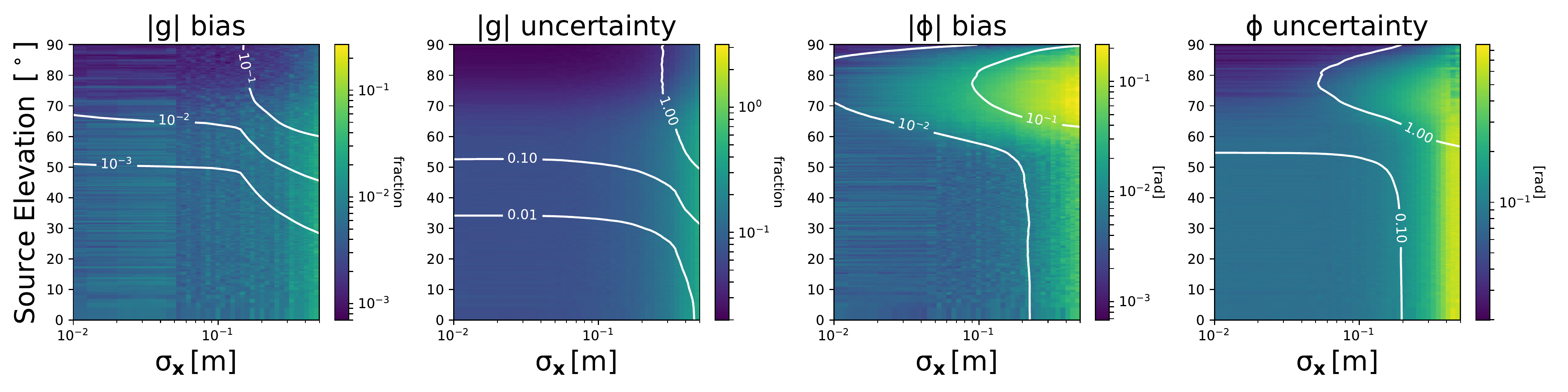}
\caption{\textit{From left to right}: the bias and uncertainty of the amplitude solutions $\lvert g \rvert$, and the bias and uncertainty of the phase solutions $\lvert \phi \rvert$ of redundant calibration on an MWA hex-like array. \textit{Top row}: we vary the position deviations $\sigma_{x}$ and source peak flux $S_{peak}$ of a source located $87^\circ$ above the horizon. \textit{Bottom row} we vary the position deviations $\sigma_{x}$ and source elevation of a $100\, \mathrm{Jy}$ source. All results are averaged over all antennas in the hex. The contour lines are the ratios between the bias and uncertainty of redundancy based and sky model based calibration, e.g. an uncertainty contour line of 2.0 indicates the uncertainty is twice as large for redundant calibration as compared to sky model based calibration.  }
\label{source_flux_source_location_and_sigma_plot}  
\end{figure*}

In our final set of redundant calibration simulations we will offset each tile in a single hex according to Gaussian distributions with mean $\mu_{\mathrm{x}} = 0$, and $\sigma_{\mathrm{x}} = 10^{-4} - 10\, \mathrm{cm}$. Now we calibrate while moving a $100\,\mathrm{Jy}$ source across the sky, and run a separate set of simulations where we fix the location of the source at $3^\circ$ off-zenith while changing its flux density instead. Figure \ref{source_flux_source_location_and_sigma_plot} shows the bias and uncertainty we derive from these simulations. To compute these from the distribution of solutions we obtain we calculate the median offset from the true gain solutions for the bias, i.e. $g = 1$ and we take the standard deviation for the uncertainty. All results are averaged over all antennas.

\begin{figure*}[htp!]
  \centering
\includegraphics[width=1\textwidth]{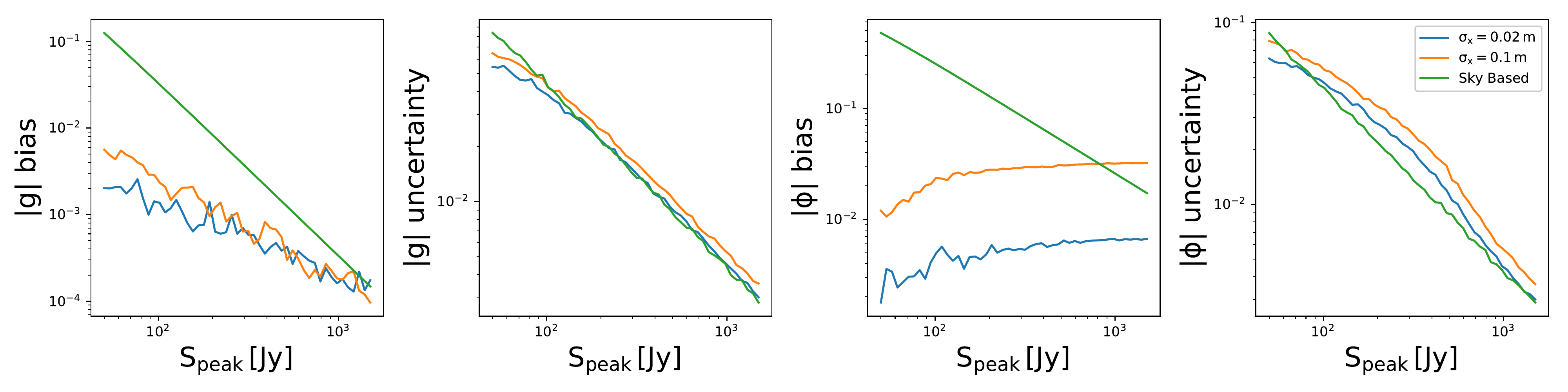}
\includegraphics[width=1\textwidth]{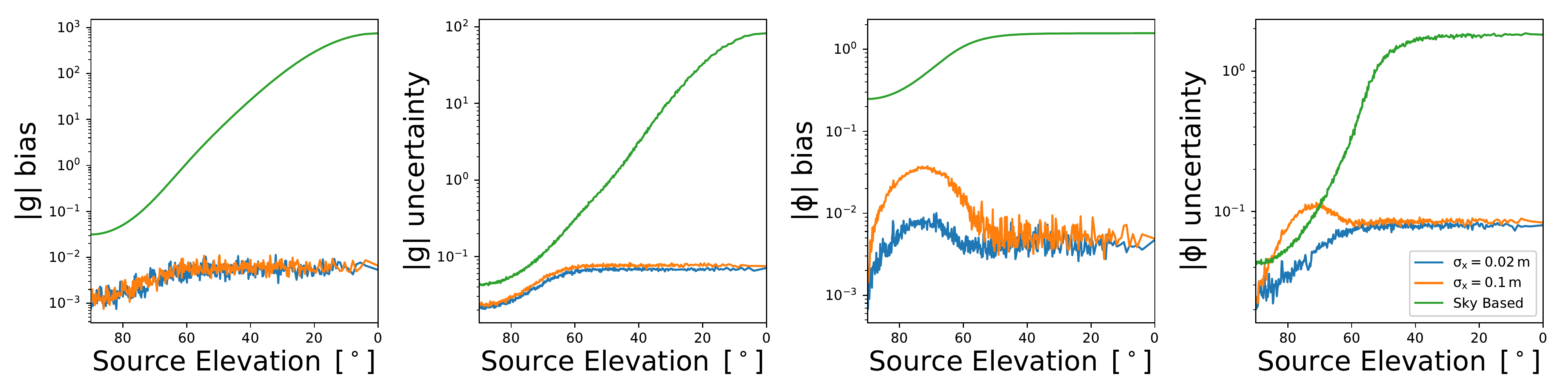}
\caption{\textit{From left to right}: the bias and uncertainty of the amplitude solutions $\lvert g \rvert$, and the bias and uncertainty of the phase solutions $\lvert \phi \rvert$ of redundant calibration on an hex array with a position precision $\sigma_x \sim 2 \, \mathrm{cm}$ . \textit{Top row}: we vary the position deviations $\sigma_{x}$ and source peak flux $S_{peak}$ of a source located $87^\circ$ above the horizon. \textit{Bottom row} we vary the position deviations $\sigma_{x}$ and source elevation of a $100\, \mathrm{Jy}$ source. All results are averaged over all antennas in the hex. The blue lines are the bias and uncertainties for an array with positional offsets $\sigma_{x} = 0.02\, \mathrm{m}$, in orange the bias and uncertainties for an array with positional offsets $\sigma_{x} = 0.1\, \mathrm{m}$, and in green the theoretical estimates for the bias and uncertainty of sky based calibration. }
\label{mwa_source_flux_and_sigma_plot}  
\end{figure*}

To compare with traditional sky-based calibration we use the bias and uncertainty derived in Section \ref{section_incomplete_sky_model}. The contour lines in Figure \ref{source_flux_source_location_and_sigma_plot} are the ratios between either the bias or uncertainty of redundant calibration and sky model calibration. To make the comparison slightly easier we take two cuts through the plots in Figure \ref{source_flux_source_location_and_sigma_plot} at an antenna position precision of $\sigma_{\mathrm{x}} = 0.02\,\mathrm{m}$ and $\sigma_{\mathrm{x}} = 0.10\, \mathrm{m}$, these cuts are shown in Figure \ref{mwa_source_flux_and_sigma_plot}.
From Figures \ref{source_flux_source_location_and_sigma_plot} and \ref{mwa_source_flux_and_sigma_plot} we can conclude the following statements. The amplitude bias depends strongly on the flux density of the primary source and its location on the sky. Redundancy-based calibration has a lower amplitude bias when the sky is dominated by a single point source, and it quickly reaches the accuracy of our implementation of the algorithm as the source moves out of the field of view.

The amplitude uncertainties of redundancy-based and sky-based calibration are comparable. For both redundant and sky model based calibration, they decrease comparably as a function of primary source flux density. However, as a function of source location redundant calibration quickly reaches the noise floor when the source is beyond the FHWM of the primary beam.

Interestingly, the phase bias increases with primary source flux density and distance of the bright source to phase centre. The bias reaches a maximum when the source is at the FWHM of the primary beam, if the source moves beyond this the bias decreases again. When the sky is dominated by a single source that is off-centre, the bias of redundant calibration becomes comparable or larger to that of sky-based calibration. The flux at which the two become comparable is dependent on the magnitude of the position offsets in the array.

The phase uncertainty depends strongly on the source flux density and its location on the sky. When the primary source brightness becomes comparable to the background sky the uncertainties of redundant become larger than that of sky model based calibration. Also note that the behaviour of the uncertainty as a function of source elevation changes for different positional precisions. When the primary source is between phase centre and the FWHM of the primary the uncertainties of redundant calibration are larger for arrays with large positional offsets. Particularly when the brightest source is at the FWHM of the primary, the uncertainties become larger than that of sky model based calibration. This increase in uncertainty can be explained by phase wrapping as discussed earlier. As the sources moves the measured visibilities phases start wrapping around $2\pi$ creating a spectrum of solutions that widens the distribution.

\section{Discussion}
\label{discussion}
The most notable results from our simulations are the results for the phase bias. Redundant calibration was proposed as an alternative to a sky model based approach because it is agnostic of the sky and therefore it does not suffer from the systematics introduced by an incomplete sky model. However, this work shows that systematics arise in a different way, because we impose the condition that our telescope is perfectly redundant. This manifests itself in systematic phase offsets in our calibration solutions because redundant calibration absorbs antenna position offsets into the calibration solutions. These phase offsets become more prominent when there is a high flux density source away from the pointing center.

\citet{Barry2016,Ewall-Wice2016,Trott2017} show that calibration on incomplete sky models causes contamination in the EoR power spectrum. Similarly redundant calibration can introduce contamination. The relative position offset changes as a function of wavelength. Therefore, the measured phase offset will therefore also vary as a function of frequency that can introduce a contamination to the EoR power spectrum.

\begin{figure*}[htp!]
  \centering
\includegraphics[width=1\textwidth]{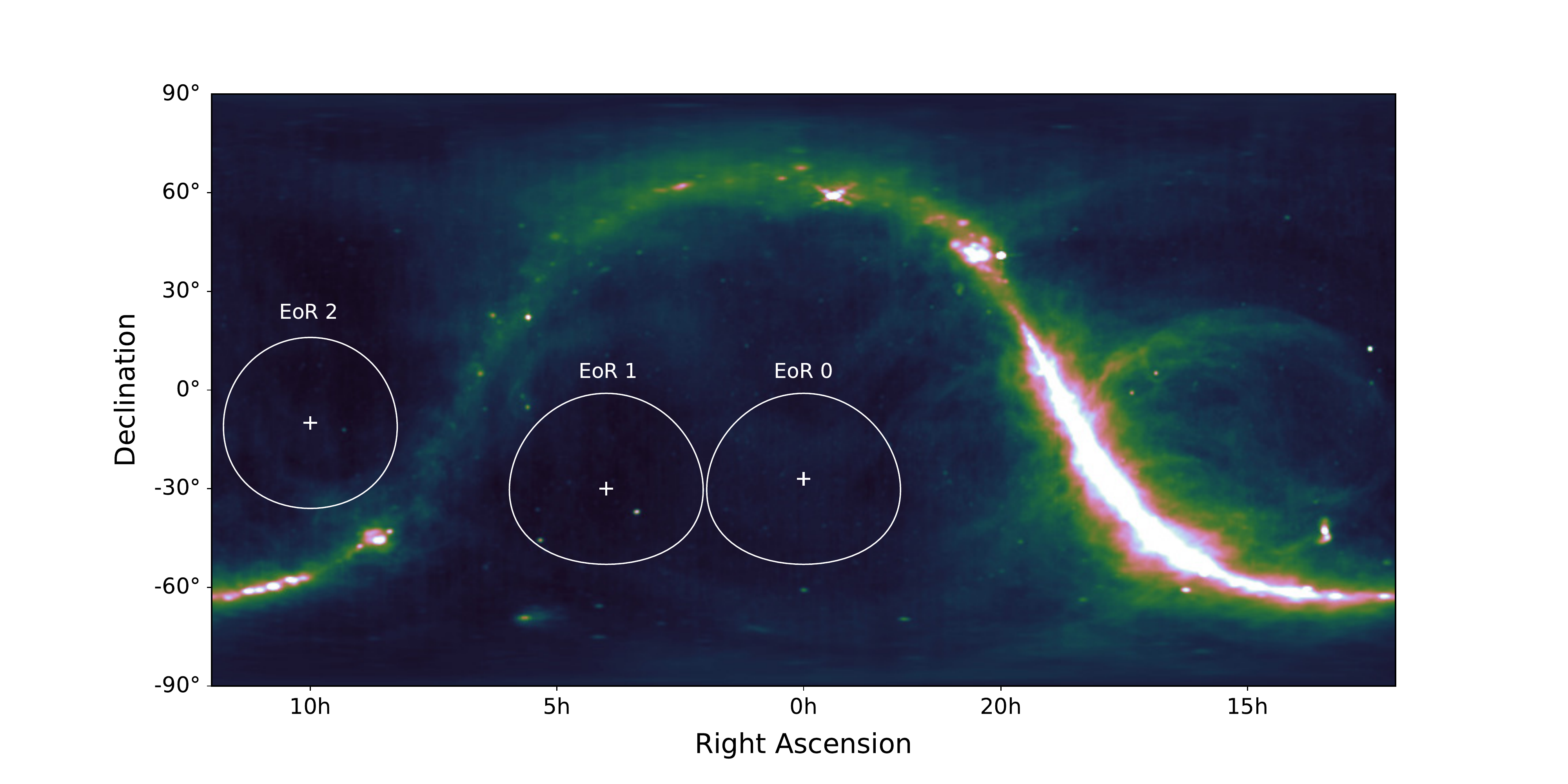}
\caption{A view of the radio sky at 408 MHz \citep{Haslam198}, the plate carr\'{e}e projection was created by \textit{Skyview}, and the location of the three MWA EoR target fields: EoR0 centered at R.A. 0h and dec. $-27^{\circ}$, EoR1 centred at R.A. 4h and dec. $-30^{\circ}$ and EoR2 centred R.A. 10h and dec. $-10^{\circ}$. The circular areas represent the FHWM of the MWA beam at 150 MHz. EoR1 and EoR2 clearly have some strong radio sources away from the point center, e.g. Fornax A and Pictor A in EoR1 and Hydra A in EoR2.}
\label{foregrounds_mwa}  
\end{figure*}

We demonstrated the influence of the sky flux distribution on the performance of redundant calibration. Figure \ref{foregrounds_mwa} shows a map of the radio sky at 408 MHz \citep{Haslam198} with the MWA EoR target fields. We can clearly see that these fields are not devoid of high flux density sources. EoR field 1 contains Fornax A and Pictor A, and EoR field 2 contains Hydra A amongst others. However, the results of the redundant calibration simulations for a single MWA hex show that position offsets at the position precision levels of the MWA are not a large source of bias and uncertainty for the phase solutions. Redundant calibration even outperforms sky based calibration on a single source. However, if a redundant array, such as HERA, has positional offsets in the order of 10~cm, careful consideration has to made on when to do redundant calibration. As demonstrated the phase bias can go up to an order of magnitude higher than that of the MWA-like array under these conditions. Fortunately for HERA the primary beam is narrower than that of the MWA, the latter suffers from significant side lobes, this and its large number of redundant baselines makes HERA somewhat robust against positional offsets \citep{Liu2010}. However, the exact trade off is still unclear.

We do note we have simplified the sky model based approach for analytic tractability. In reality a sky model will contain more than 1 calibration source, therefore the bias and uncertainty for a sky based approach will certainly be lower than presented here. But as a general lesson we can conclude that for redundant calibration it is preferable to have strong sources like those present in the EoR1 and EoR2 fields either at the pointing center or at the edge of the beam. EoR field 0 would therefore be an excellent field for redundant calibration.

In this work we have not considered the differences in the antenna response of different antenna tiles. Work by \citet{Line2018}) shows that the tile beam differences are on the order of 10\%. This poses most likely the largest hurdle for redundant calibration. Studying the effect of these beam differences and how it impacts the redundancy of the MWA hexes and other radio telescopes is therefore crucial to understand the limitations of redundant calibration in realistic telescopes.

This is also where the true strength of sky based calibration methods comes into play. Because redundant calibration relies on the assumption that each antenna observes the same radio sky, it also is unable to solve for direction dependent effects introduced by different antenna responses and ionospheric distortions. The field of direction dependent calibration faced quite a number of challenges, e.g. diffuse emission detected by shorter baselines, solving for enough different directions to capture variations in the ionosphere or the primary beam responses, optimizing the calibration time scale to reduce computational load, and the observed curvature of the sky (w-correction) due to the wide FoVs of these new arrays. A large effort has gone into solving these issues, e.g. \texttt{SAGEcal} resolved the computational load of solving for a large number of directions by using the SAGE algorithm rather than traditional least squares optimization \citep{Yatawatta2009}, facet calibration divides the sky in facets to reduce the number of parameters to solve for simultaneously \citep{Weeren2016}, \texttt{RTS} employs the MWA's uv-coverage to perform snapshot imaging tackling the w-term problem \citep{Mitchell2008}, with the diffuse emission of the Milky Way remaining as a major challenge. We have only mentioned a few implementations available as each science case has its own goal accompanied with its own implementation of sky based calibration. But the result of this large effort are impressive high-fidelity images, required to either study foreground sources or to subtract them. The latter being the goal for the EoR experiment.  
Solving for these higher-order calibration features is, however, out of reach for redundant calibration. Furthermore, redundant calibration does not truly escape the need for a sky model, because the degenerate parameters in Equation \ref{redundant_calibration_degeneracy} need to be constrained by external information, i.e sky-based calibration. \citet{Li2018} directly compare redundant calibration using \texttt{OMNICAL} and sky model calibration using \texttt{FHD} and find that they perform similarly on data from real MWA EoR observations, that include the position offsets and tile beam differences. \citet{Li2018} also investigate the complimentary nature of the two different calibration techniques and find that combining the two methods improves the sensitivity to the EoR power spectrum, demonstrating that ``hybrid approaches'' are the best way forward. However, this final step can also introduce errors with spectral structure due to an incomplete sky model.
Redundant calibration is, at best, a way to add another constraint for a first order calibration step. Higher order effects require a pure sky based calibration that include direction dependent effects. Nevertheless, redundant calibration can still add useful information if carefully applied. \citet{Sievers2017} propose a calibration algorithm that sits in the middle ground between agnostic redundant calibration and pure sky model based calibration maximally using the information in the data of a generic radio telescopes. This methodology seems to be a promising avenue for the MWA and the future SKA.

\section{Conclusion and Future Work}
\label{conclusion}
In this work we use a rudimentary implementation of the \texttt{logcal} and \texttt{lincal} algorithm to understand the fundamental limitations of redundant calibration on nearly redundant telescopes. We simulate redundant calibration under different radio sky conditions and find that the phase solutions are systematically impacted by position offsets in a redundant telescope. Based on our simulations we conclude the following key statement: redundant calibration performs best when strong radio sources are either at field center or at the edge of the primary beam. 
We also compare redundant calibration to sky model based calibration and find that for the MWA redundant calibration of the redundant hexes performs better than a sky based approach. However,  we require further work to understand the impact of non-redundancies introduced by differences in tile beam responses that may be of larger concern to the MWA EoR experiment. Moreover we also assumed that the sky model consists of only a single source. More work would be required to understand the completeness threshold above which a sky based approach truly outperforms redundant calibration on a nearly redundant array. Finally, to optimally calibrate our radio telescopes we require a hybrid approach that bridges the gap between redundancy based and sky based calibration, and we see this as the way forward in the calibration of the current and next generation of radio telescopes.  

\acknowledgments
RCJ thanks Ruby Byrne for useful discussions, and Wenyang Li, Bart Pindor and Christopher Jordan for constructive feedback on this manuscript. This work was supported by the Centre for All Sky Astrophysics in 3 Dimensions (ASTRO 3D), an Australian Research Council Centre of Excellence, funded by grant CE170100013.  This research has made use of NASA's Astrophysics Data System. We acknowledge the International Centre for Radio Astronomy Research (ICRAR), a Joint Venture of Curtin University and The University of Western Australia, funded by the Western Australian State government. This scientific work makes use of the Murchison Radio-astronomy Observatory, operated by CSIRO. We acknowledge the Wajarri Yamatji people as the traditional owners of the Observatory site. Support for the operation of the MWA is provided by the Australian Government (NCRIS), under a contract to Curtin University administered by Astronomy Australia Limited. 



\appendix
\section{Multi Frequency Implementation}
\label{multi_frequency}
\begin{figure*}[htp]
  \centering
\includegraphics[width=1\textwidth]{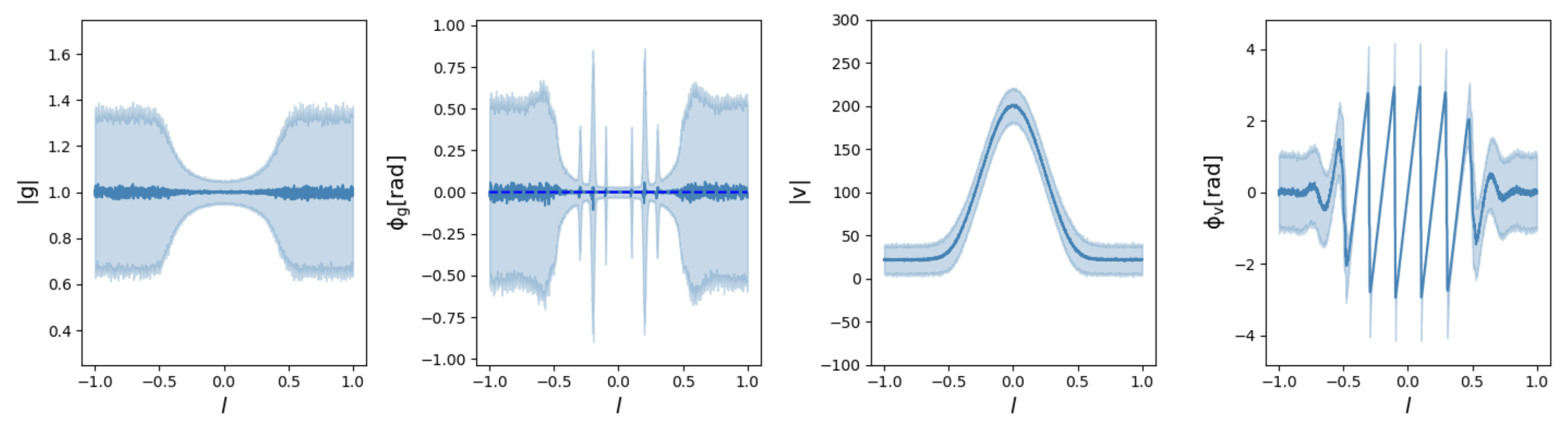}
\caption {From left to right: the amplitude gain, the phase gain solutions for a single antenna in an ideal 5 element interferometer, the visibility amplitude and the visibility phase for one of the two redundant groups as a function of strong source position $l$. The dark blue line represents the mean of the solutions, the shaded blue area indicates the 1-sigma solutions variance. The amplitude solution variance inversely follows the shape of the beam, i.e. the variance increases when the beam response decreases. The mean of the phase solutions generally fluctuates around $\phi =0$ and deviates along with the variance at so-called phase wrapping points, which are further explained in the text. The additional position offset, which causes a phase offset from the ideal redundant phase, is absorbed into the solutions causing a slope.}
\label{multichannel_5_element}  
\end{figure*}

Redundant calibration is typically presented on a channel-by-channel basis in contrast with standard model based calibration schemes that operate over a range of frequencies. The multi frequency approach uses all of the information available in other frequency channels. In this section we will discuss the multi-frequency implementation of \texttt{logcal}, allowing it to benefit from the multi-frequency information available in radio interferometry data. Another motivation is that a multi-frequency implementation has the prospect of resolving the phase wrapping problem. Earlier, we described that phase wrapping occurs when a specific baseline $u$ observes a source on a specific coordinate $l = 1/2u$, the phase becomes ill-defined at this point. That same baseline should measure a defined phase when observing that same source at a different frequency. This property has motivated us to extend the classical redundant calibration framework by solving for the gain and visibilities simultaneously at different frequency channels. We will assume that the gain is the same at those frequencies, as a first order approximation, but the visibilities are different. \\
We extend Equation \ref{matrix_equation}, by stacking the measurement vectors at different frequencies and adding the visibilities from the additional frequency channels. We then construct the matrix accordingly by realizing that $\mathbf{A}$ can be split into two components:

\begin{equation}
\begin{aligned}
\mathbf{A}_{g} = 
\begin{pmatrix}
-0 & 1 & 0 & 0 & 0\\
0 & -1 & 1 & 0 & 0\\
0 & 0 & -1 & 1 & 0\\
0 & 0 & 0 & -1 & 1\\
-1 & 0 & 1 & 0 & 0\\
0 & -1 & 0 & 1 & 0\\
0 & 0 & -1 & 0 & 0\\
\end{pmatrix}
&\mathbf{A}_{v} = 
\begin{pmatrix}
1 & 0 \\
1 & 0 \\
1 & 0 \\
1 & 0 \\
0 & 1 \\
0 & 1 \\
0 & 1 \\
\end{pmatrix}
\end{aligned}
,
\label{gain_matrix}
\end{equation}

where $\mathbf{A}_{g}$ maps the gains onto the measurements and $\mathbf{A}_{v}$ maps the visibilities onto the measurements. We can construct a multi frequency matrix combining $\mathbf{A}_{g}$ and $\mathbf{A}_{v}$, e.g. for a two-channel solutions estimation.
\begin{equation}
\begin{aligned}
\mathbf{A} = 
\begin{pmatrix}
\mathbf{A}_{g} & \mathbf{A}_{v} & \mathbf{0}\\
\mathbf{A}_{g} & \mathbf{0} & \mathbf{A}_{v}\\
\end{pmatrix}
\end{aligned}
\label{multi_channel_matrix}
\end{equation}
Using this extended version of redundant calibration we return to the 5-element interferometer, while varying the number of channels involved to calibrate our antennas. Figure \ref{multichannel_5_element} shows the results when we attempt to calibrate 5 antennas using 2 frequency channels spaced around the actual frequency channel we are interested in. We observe a significant change in the structure of the variance of the phase solutions. The peaks around the phase wrapping points have severely decreased in width. However, it has not resolved the the phase wrapping point, which was the aim of this multi frequency implementation.

To understand why a multi frequency extension of redundant calibration does not solve the problems immediately we have to return to Equation \ref{matrix_equation}. The matrix $[\mathbf{A}^{T}\mathbf{A}]^{-1}\mathbf{A}^{T}$ mixes the phase wrapping and the non-phase wrapping channels into the calibration solutions. Although adding extra frequency channels does adds another set of constraints to the calibration solutions, the solutions do no escape the impact of the phase wrapping channel. Adding more channels therefore would decrease variance, but not resolve a phase wrapping point. To really resolve phase wrapping in redundant calibration we require the inclusion of knowledge of the sky.

\bibliographystyle{aasjournal}

\bibliography{0catalogue}

\begin{thebibliography}{}
\expandafter\ifx\csname natexlab\endcsname\relax\def\natexlab#1{#1}\fi
\providecommand{\url}[1]{\href{#1}{#1}}

\bibitem[{Ali {et~al.}(2015)Ali, Parsons, Zheng, Pober, Liu, Aguirre, Bradley,
  Bernardi, Carilli, Cheng, \& et~al.}]{Ali2015}
Ali, Z.~S., Parsons, A.~R., Zheng, H., {et~al.} 2015, The Astrophysical
  Journal, 809, 61.
\newblock \url{http://dx.doi.org/10.1088/0004-637X/809/1/61}

\bibitem[{{Barry} {et~al.}(2016){Barry}, {Hazelton}, {Sullivan}, {Morales}, \&
  {Pober}}]{Barry2016}
{Barry}, N., {Hazelton}, B., {Sullivan}, I., {Morales}, M.~F., \& {Pober},
  J.~C. 2016, \mnras, 461, 3135

\bibitem[{Beardsley {et~al.}(2016)Beardsley, Hazelton, Sullivan, Carroll,
  Barry, Rahimi, Pindor, Trott, Line, Jacobs, Morales, Pober, Bernardi, Bowman,
  Busch, Briggs, Cappallo, Corey, de~Oliveira-Costa, Dillon, Emrich,
  Ewall-Wice, Feng, Gaensler, Goeke, Greenhill, Hewitt, Hurley-Walker,
  Johnston-Hollitt, Kaplan, Kasper, Kim, Kratzenberg, Lenc, Loeb, Lonsdale,
  Lynch, McKinley, McWhirter, Mitchell, Morgan, Neben, Thyagarajan, Oberoi,
  Offringa, Ord, Paul, Prabu, Procopio, Riding, Rogers, Roshi, Shankar, Sethi,
  Srivani, Subrahmanyan, Tegmark, Tingay, Waterson, Wayth, Webster, Whitney,
  Williams, Williams, Wu, \& Wyithe}]{Beardsley2016}
Beardsley, A.~P., Hazelton, B.~J., Sullivan, I.~S., {et~al.} 2016,
  arXiv:1608.06281

\bibitem[{Bowman {et~al.}(2006)Bowman, Morales, \& Hewitt}]{Bowman2006}
Bowman, J.~D., Morales, M.~F., \& Hewitt, J.~N. 2006, ApJ, 638, 20

\bibitem[{{Datta} {et~al.}(2009){Datta}, {Bhatnagar}, \& {Carilli}}]{Datta2009}
{Datta}, A., {Bhatnagar}, S., \& {Carilli}, C.~L. 2009, \apj, 703, 1851

\bibitem[{{Datta} {et~al.}(2010){Datta}, {Bowman}, \& {Carilli}}]{Datta2010}
{Datta}, A., {Bowman}, J.~D., \& {Carilli}, C.~L. 2010, \apj, 724, 526

\bibitem[{{Dillon} \& {Parsons}(2016)}]{Dillon2016}
{Dillon}, J.~S., \& {Parsons}, A.~R. 2016, \apj, 826, 181

\bibitem[{Ewall-Wice {et~al.}(2016)Ewall-Wice, Dillon, Liu, \&
  Hewitt}]{Ewall-Wice2016}
Ewall-Wice, A., Dillon, J.~S., Liu, A., \& Hewitt, J. 2016, 1610.02689.
\newblock \url{https://arxiv.org/abs/1610.02689}

\bibitem[{Franzen {et~al.}(2016)Franzen, Jackson, Offringa, Ekers, Wayth,
  Bernardi, Bowman, Briggs, Cappallo, Deshpande, Gaensler, Greenhill, Hazelton,
  Johnston-Hollitt, Kaplan, Lonsdale, McWhirter, Mitchell, Morales, Morgan,
  Morgan, Oberoi, Ord, Prabu, Seymour, Shankar, Srivani, Subrahmanyan, Tingay,
  Trott, Webster, Williams, \& Williams}]{Franzen2016}
Franzen, T. M.~O., Jackson, C.~A., Offringa, A.~R., {et~al.} 2016, \mnras, 459,
  3314

\bibitem[{Furlanetto {et~al.}(2006)Furlanetto, Oh, \& Briggs}]{Furlanetto2006}
Furlanetto, S., Oh, S.~P., \& Briggs, F. 2006, Phys.Rept., 433, 181

\bibitem[{{Furlanetto}(2016)}]{Furlanetto2016}
{Furlanetto}, S.~R. 2016, in Astrophysics and Space Science Library, Vol. 423,
  Understanding the Epoch of Cosmic Reionization: Challenges and Progress, ed.
  A.~{Mesinger}, 247

\bibitem[{{Gervasi} {et~al.}(2008){Gervasi}, {Tartari}, {Zannoni}, {Boella}, \&
  {Sironi}}]{Gervasi2008}
{Gervasi}, M., {Tartari}, A., {Zannoni}, M., {Boella}, G., \& {Sironi}, G.
  2008, \apj, 682, 223

\bibitem[{Grobler {et~al.}(2014)Grobler, Nunhokee, Smirnov, van Zyl, \&
  de~Bruyn}]{Grobler2014}
Grobler, T.~L., Nunhokee, C.~D., Smirnov, O.~M., van Zyl, A.~J., \& de~Bruyn,
  A.~G. 2014, \mnras, 439, 4030

\bibitem[{{Haslam} {et~al.}(1982){Haslam}, {Salter}, {Stoffel}, \&
  {Wilson}}]{Haslam198}
{Haslam}, C.~G.~T., {Salter}, C.~J., {Stoffel}, H., \& {Wilson}, W.~E. 1982,
  Astronomy and Astrophysics Supplement Series, 47, 1

\bibitem[{Intema {et~al.}(2009)Intema, van~der Tol, Cotton, Cohen, van Bemmel,
  \& R{\"o}ttgering}]{Intema2009}
Intema, H.~T., van~der Tol, S., Cotton, W.~D., {et~al.} 2009, \aap, 501, 1185.
\newblock \url{http://adsabs.harvard.edu/abs/2009A%26A...501.1185I}

\bibitem[{{Intema} {et~al.}(2011){Intema}, {van Weeren}, {R{\"o}ttgering}, \&
  {Lal}}]{Intema2011}
{Intema}, H.~T., {van Weeren}, R.~J., {R{\"o}ttgering}, H.~J.~A., \& {Lal},
  D.~V. 2011, \aap, 535, arXiv:1109.5906

\bibitem[{Jelic {et~al.}(2008)Jelic, Zaroubi, Labropoulos, Thomas, Bernardi,
  {et~al.}}]{Jelic2008}
Jelic, V., Zaroubi, S., Labropoulos, P., {et~al.} 2008,
  Mon.Not.Roy.Astron.Soc., 389, 1319

\bibitem[{Joseph(2018)}]{Joseph2018a}
Joseph, R. 2018, ronniyjoseph/SCAR: First Release, , ,
  doi:10.5281/zenodo.1463587.
\newblock \url{https://doi.org/10.5281/zenodo.1463587}

\bibitem[{Kay(1993)}]{Kay1993}
Kay, S.~M. 1993, Fundamentals of Statistical Signal Processing: Estimation
  Theory (Upper Saddle River, NJ, USA: Prentice-Hall, Inc.)

\bibitem[{Kazemi {et~al.}(2011)Kazemi, Yatawatta, Zaroubi, Lampropoulos,
  de~Bruyn, Koopmans, \& Noordam}]{Kazemi2011}
Kazemi, S., Yatawatta, S., Zaroubi, S., {et~al.} 2011, \mnras, 414, 1656.
\newblock \url{http://adsabs.harvard.edu/abs/2011MNRAS.414.1656K}

\bibitem[{Li {et~al.}(2018)Li, Pober, Hazelton, Barry, Morales, Sullivan,
  Parsons, Ali, Dillon, Beardsley, Bowman, Briggs, Byrne, Carroll, Crosse,
  Emrich, Ewall-Wice, Feng, Franzen, Hewitt, Horsley, Jacobs, Johnston-Hollitt,
  Jordan, Joseph, Kaplan, Kenney, Kim, Kittiwisit, Lanman, Line, McKinley,
  Mitchell, Murray, Neben, Offringa, Pallot, Paul, Pindor, Procopio, Rahimi,
  Riding, Sethi, Shankar, Steele, Subrahmanian, Tegmark, Thyagarajan, Tingay,
  Trott, Walker, Wayth, Webster, Williams, Wu, \& Wyithe}]{Li2018}
Li, W., Pober, J.~C., Hazelton, B.~J., {et~al.} 2018, 1807.05312v1

\bibitem[{Line {et~al.}(2018)Line, McKinley, Rasti, Bhardwaj, Wayth, Webster,
  Ung, Emrich, Horsley, Beardsley, Crosse, Franzen, Gaensler, Johnston-Hollitt,
  Kaplan, Kenney, Morales, Pallot, Steele, Tingay, Trott, Walker, Williams, \&
  Wu}]{Line2018}
Line, J. L.~B., McKinley, B., Rasti, J., {et~al.} 2018, 1808.04516

\bibitem[{Liu {et~al.}(2010)Liu, Tegmark, Morrison, Lutomirski, \&
  Zaldarriaga}]{Liu2010}
Liu, A., Tegmark, M., Morrison, S., Lutomirski, A., \& Zaldarriaga, M. 2010,
  Monthly Notices of the Royal Astronomical Society, 408, 1029–1050.
\newblock \url{http://dx.doi.org/10.1111/j.1365-2966.2010.17174.x}

\bibitem[{McQuinn(2015)}]{McQuinn2015}
McQuinn, M. 2015, 1512.00086

\bibitem[{{Mitchell} {et~al.}(2008){Mitchell}, {Greenhill}, {Wayth}, {Sault},
  {Lonsdale}, {Cappallo}, {Morales}, \& {Ord}}]{Mitchell2008}
{Mitchell}, D.~A., {Greenhill}, L.~J., {Wayth}, R.~B., {et~al.} 2008, IEEE
  Journal of Selected Topics in Signal Processing, 2, 707

\bibitem[{{Morales} {et~al.}(2006){Morales}, {Bowman}, \&
  {Hewitt}}]{Morales2006}
{Morales}, M.~F., {Bowman}, J.~D., \& {Hewitt}, J.~N. 2006, \apj, 648, 767

\bibitem[{{Morales} {et~al.}(2012){Morales}, {Hazelton}, {Sullivan}, \&
  {Beardsley}}]{Morales2012}
{Morales}, M.~F., {Hazelton}, B., {Sullivan}, I., \& {Beardsley}, A. 2012,
  \apj, 752, 137

\bibitem[{Morales \& Wyithe(2010)}]{Morales2010}
Morales, M.~F., \& Wyithe, J. S.~B. 2010, Ann.Rev.Astron.Astrophys., 48, 127

\bibitem[{Murray(2018)}]{Murray2018}
Murray, S.~G. 2018, 3, 850

\bibitem[{Murray {et~al.}(2017)Murray, Trott, \& Jordan}]{Murray2017}
Murray, S.~G., Trott, C.~M., \& Jordan, C.~H. 2017, 1706.10033

\bibitem[{Noorishad {et~al.}(2012)Noorishad, Wijnholds, van Ardenne, \& van~der
  Hulst}]{Noorishad2012}
Noorishad, P., Wijnholds, S.~J., van Ardenne, A., \& van~der Hulst, J.~M. 2012,
  Astronomy \& Astrophysics, 545, A108.
\newblock \url{http://dx.doi.org/10.1051/0004-6361/201219087}

\bibitem[{{Parsons} {et~al.}(2010){Parsons}, {Backer}, {Foster}, {Wright},
  {Bradley}, {Gugliucci}, {Parashare}, {Benoit}, {Aguirre}, {Jacobs},
  {Carilli}, {Herne}, {Lynch}, {Manley}, \& {Werthimer}}]{Parsons2010}
{Parsons}, A.~R., {Backer}, D.~C., {Foster}, G.~S., {et~al.} 2010, \aj, 139,
  1468

\bibitem[{{Patil} {et~al.}(2017){Patil}, {Yatawatta}, {Koopmans}, {de Bruyn},
  {Brentjens}, {Zaroubi}, {Asad}, {Hatef}, {Jeli{\'c}}, {Mevius}, {Offringa},
  {Pandey}, {Vedantham}, {Abdalla}, {Brouw}, {Chapman}, {Ciardi}, {Gehlot},
  {Ghosh}, {Harker}, {Iliev}, {Kakiichi}, {Majumdar}, {Mellema}, {Silva},
  {Schaye}, {Vrbanec}, \& {Wijnholds}}]{Patil2017}
{Patil}, A.~H., {Yatawatta}, S., {Koopmans}, L.~V.~E., {et~al.} 2017, \apj,
  838, 65

\bibitem[{{Pritchard} \& {Loeb}(2008)}]{Pritchard2008}
{Pritchard}, J., \& {Loeb}, A. 2008, Physical review D, 78, 103511

\bibitem[{{Pritchard} \& {Loeb}(2012)}]{Pritchard2012}
{Pritchard}, J.~R., \& {Loeb}, A. 2012, Reports on Progress in Physics, 75,
  086901

\bibitem[{Rau {et~al.}(2009)Rau, Bhatnagar, Voronkov, \& Cornwell}]{Rau2009}
Rau, U., Bhatnagar, S., Voronkov, M., \& Cornwell, T. 2009, Proceedings of the
  IEEE, 97, 1472–1481.
\newblock \url{http://dx.doi.org/10.1109/JPROC.2009.2014853}

\bibitem[{Salvini \& Wijnholds(2014)}]{Salvini2014}
Salvini, S., \& Wijnholds, S.~J. 2014, \aap, 571, A97.
\newblock \url{http://adsabs.harvard.edu/abs/2014A%26A...571A..97S}

\bibitem[{Sievers(2017)}]{Sievers2017}
Sievers, J.~L. 2017, 1701.01860

\bibitem[{Sullivan {et~al.}(2012)Sullivan, Morales, Hazelton, Arcus, Barnes,
  Bernardi, Briggs, Bowman, Bunton, Cappallo, Corey, Deshpande, deSouza,
  Emrich, Gaensler, Goeke, Greenhill, Herne, Hewitt, Johnston-Hollitt, Kaplan,
  Kasper, Kincaid, Koenig, Kratzenberg, Lonsdale, Lynch, McWhirter, Mitchell,
  Morgan, Oberoi, Ord, Pathikulangara, Prabu, Remillard, Rogers, Roshi, Salah,
  Sault, Udaya~Shankar, Srivani, Stevens, Subrahmanyan, Tingay, Wayth,
  Waterson, Webster, Whitney, Williams, Williams, \& Wyithe}]{Sullivan2012}
Sullivan, I.~S., Morales, M.~F., Hazelton, B.~J., {et~al.} 2012, \apj, 759, 17.
\newblock \url{http://adsabs.harvard.edu/abs/2012ApJ...759...17S}

\bibitem[{Tingay {et~al.}(2013)Tingay, Goeke, Bowman, Emrich, Ord, Mitchell,
  Morales, Booler, Crosse, Wayth, \& et~al.}]{Tingay2013}
Tingay, S.~J., Goeke, R., Bowman, J.~D., {et~al.} 2013, Publications of the
  Astronomical Society of Australia, 30, doi:10.1017/pasa.2012.007.
\newblock \url{http://dx.doi.org/10.1017/pasa.2012.007}

\bibitem[{{Trott} \& {Wayth}(2017)}]{Trott2017}
{Trott}, C.~M., \& {Wayth}, R.~B. 2017, \pasa, 34, e061

\bibitem[{Trott {et~al.}(2012)Trott, Wayth, \& Tingay}]{Trott2012}
Trott, C.~M., Wayth, R.~B., \& Tingay, S.~J. 2012, The Astrophysical Journal,
  757, 101

\bibitem[{{Trott} {et~al.}(2016){Trott}, {Pindor}, {Procopio}, {Wayth},
  {Mitchell}, {McKinley}, {Tingay}, {Barry}, {Beardsley}, {Bernardi}, {Bowman},
  {Briggs}, {Cappallo}, {Carroll}, {de Oliveira-Costa}, {Dillon}, {Ewall-Wice},
  {Feng}, {Greenhill}, {Hazelton}, {Hewitt}, {Hurley-Walker},
  {Johnston-Hollitt}, {Jacobs}, {Kaplan}, {Kim}, {Lenc}, {Line}, {Loeb},
  {Lonsdale}, {Morales}, {Morgan}, {Neben}, {Thyagarajan}, {Oberoi},
  {Offringa}, {Ord}, {Paul}, {Pober}, {Prabu}, {Riding}, {Udaya Shankar},
  {Sethi}, {Srivani}, {Subrahmanyan}, {Sullivan}, {Tegmark}, {Webster},
  {Williams}, {Williams}, {Wu}, \& {Wyithe}}]{Trott2016}
{Trott}, C.~M., {Pindor}, B., {Procopio}, P., {et~al.} 2016, \apj, 818, 139

\bibitem[{{van Haarlem} {et~al.}(2013){van Haarlem}, {Wise}, {Gunst}, {Heald},
  {McKean}, {Hessels}, {de Bruyn}, {Nijboer}, {Swinbank}, {Fallows},
  {Brentjens}, {Nelles}, {Beck}, {Falcke}, {Fender}, {H{\"o}randel},
  {Koopmans}, {Mann}, {Miley}, {R{\"o}ttgering}, {Stappers}, {Wijers},
  {Zaroubi}, {van den Akker}, {Alexov}, {Anderson}, {Anderson}, {van Ardenne},
  {Arts}, {Asgekar}, {Avruch}, {Batejat}, {B{\"a}hren}, {Bell}, {Bell}, {van
  Bemmel}, {Bennema}, {Bentum}, {Bernardi}, {Best}, {B{\^i}rzan}, {Bonafede},
  {Boonstra}, {Braun}, {Bregman}, {Breitling}, {van de Brink}, {Broderick},
  {Broekema}, {Brouw}, {Br{\"u}ggen}, {Butcher}, {van Cappellen}, {Ciardi},
  {Coenen}, {Conway}, {Coolen}, {Corstanje}, {Damstra}, {Davies}, {Deller},
  {Dettmar}, {van Diepen}, {Dijkstra}, {Donker}, {Doorduin}, {Dromer}, {Drost},
  {van Duin}, {Eisl{\"o}ffel}, {van Enst}, {Ferrari}, {Frieswijk}, {Gankema},
  {Garrett}, {de Gasperin}, {Gerbers}, {de Geus}, {Grie{\ss}meier}, {Grit},
  {Gruppen}, {Hamaker}, {Hassall}, {Hoeft}, {Holties}, {Horneffer}, {van der
  Horst}, {van Houwelingen}, {Huijgen}, {Iacobelli}, {Intema}, {Jackson},
  {Jelic}, {de Jong}, {Juette}, {Kant}, {Karastergiou}, {Koers}, {Kollen},
  {Kondratiev}, {Kooistra}, {Koopman}, {Koster}, {Kuniyoshi}, {Kramer},
  {Kuper}, {Lambropoulos}, {Law}, {van Leeuwen}, {Lemaitre}, {Loose}, {Maat},
  {Macario}, {Markoff}, {Masters}, {McFadden}, {McKay-Bukowski}, {Meijering},
  {Meulman}, {Mevius}, {Middelberg}, {Millenaar}, {Miller-Jones}, {Mohan},
  {Mol}, {Morawietz}, {Morganti}, {Mulcahy}, {Mulder}, {Munk}, {Nieuwenhuis},
  {van Nieuwpoort}, {Noordam}, {Norden}, {Noutsos}, {Offringa}, {Olofsson},
  {Omar}, {Orr{\'u}}, {Overeem}, {Paas}, {Pandey-Pommier}, {Pandey}, {Pizzo},
  {Polatidis}, {Rafferty}, {Rawlings}, {Reich}, {de Reijer}, {Reitsma},
  {Renting}, {Riemers}, {Rol}, {Romein}, {Roosjen}, {Ruiter}, {Scaife}, {van
  der Schaaf}, {Scheers}, {Schellart}, {Schoenmakers}, {Schoonderbeek},
  {Serylak}, {Shulevski}, {Sluman}, {Smirnov}, {Sobey}, {Spreeuw}, {Steinmetz},
  {Sterks}, {Stiepel}, {Stuurwold}, {Tagger}, {Tang}, {Tasse}, {Thomas},
  {Thoudam}, {Toribio}, {van der Tol}, {Usov}, {van Veelen}, {van der Veen},
  {ter Veen}, {Verbiest}, {Vermeulen}, {Vermaas}, {Vocks}, {Vogt}, {de Vos},
  {van der Wal}, {van Weeren}, {Weggemans}, {Weltevrede}, {White}, {Wijnholds},
  {Wilhelmsson}, {Wucknitz}, {Yatawatta}, {Zarka}, {Zensus}, \& {van
  Zwieten}}]{vanHaarlem2013}
{van Haarlem}, M.~P., {Wise}, M.~W., {Gunst}, A.~W., {et~al.} 2013, \aap, 556,
  A2

\bibitem[{van Weeren {et~al.}(2016)van Weeren, Williams, Hardcastle, Shimwell,
  Rafferty, Sabater, Heald, Sridhar, Dijkema, Brunetti, Br{\"u}ggen,
  Andrade-Santos, Ogrean, R{\"o}ttgering, Dawson, Forman, de~Gasperin, Jones,
  Miley, Rudnick, Sarazin, Bonafede, Best, B{\^i}rzan, Cassano, Chy{\.z}y,
  Croston, Ensslin, Ferrari, Hoeft, Horellou, Jarvis, Kraft, Mevius, Intema,
  Murray, Orr{\'u}, Pizzo, Simionescu, Stroe, van~der Tol, \&
  White}]{Weeren2016}
van Weeren, R.~J., Williams, W.~L., Hardcastle, M.~J., {et~al.} 2016, \apjs,
  223, 2.
\newblock \url{http://adsabs.harvard.edu/abs/2016ApJS..223....2V}

\bibitem[{{Vedantham} {et~al.}(2012){Vedantham}, {Udaya Shankar}, \&
  {Subrahmanyan}}]{Vedantham2012}
{Vedantham}, H., {Udaya Shankar}, N., \& {Subrahmanyan}, R. 2012, \apj, 745,
  176

\bibitem[{Wieringa(1992)}]{Wieringa1992}
Wieringa, M.~H. 1992, Experimental Astronomy, 2, 203–225.
\newblock \url{http://dx.doi.org/10.1007/BF00420576}

\bibitem[{{Wijnholds} {et~al.}(2010){Wijnholds}, {van der Tol}, {Nijboer}, \&
  {van der Veen}}]{Wijnholds2010}
{Wijnholds}, S., {van der Tol}, S., {Nijboer}, R., \& {van der Veen}, A.-J.
  2010, IEEE Signal Processing Magazine, 27, 30

\bibitem[{Wijnholds {et~al.}(2016)Wijnholds, Grobler, \&
  Smirnov}]{Wijnholds2016}
Wijnholds, S.~J., Grobler, T.~L., \& Smirnov, O.~M. 2016, \mnras, 457, 2331

\bibitem[{{Williams} {et~al.}(2016){Williams}, {van Weeren}, {R{\"o}ttgering},
  {Best}, {Dijkema}, {de Gasperin}, {Hardcastle}, {Heald}, {Prandoni},
  {Sabater}, {Shimwell}, {Tasse}, {van Bemmel}, {Br{\"u}ggen}, {Brunetti},
  {Conway}, {En{\ss}lin}, {Engels}, {Falcke}, {Ferrari}, {Haverkorn},
  {Jackson}, {Jarvis}, {Kapi{\'n}ska}, {Mahony}, {Miley}, {Morabito},
  {Morganti}, {Orr{\'u}}, {Retana-Montenegro}, {Sridhar}, {Toribio}, {White},
  {Wise}, \& {Zwart}}]{Williams2016}
{Williams}, W.~L., {van Weeren}, R.~J., {R{\"o}ttgering}, H.~J.~A., {et~al.}
  2016, \mnras, 460, 2385

\bibitem[{Yatawatta {et~al.}(2009)Yatawatta, Zaroubi, de~Bruyn, Koopmans, \&
  Noordam}]{Yatawatta2009}
Yatawatta, S., Zaroubi, S., de~Bruyn, G., Koopmans, L., \& Noordam, J. 2009, in
  Proc. IEEE 13th Digital Signal Processing Workshop and 5th IEEE Signal
  Processing Education Workshop, 150--155

\bibitem[{Zheng {et~al.}(2014)Zheng, Tegmark, Buza, Dillon, Gharibyan, Hickish,
  Kunz, Liu, Losh, Lutomirski, \& et~al.}]{Zheng2014}
Zheng, H., Tegmark, M., Buza, V., {et~al.} 2014, Monthly Notices of the Royal
  Astronomical Society, 445, 1084–1103.
\newblock \url{http://dx.doi.org/10.1093/mnras/stu1773}

\end{thebibliography}

\end{document}